\documentclass[10pt,a4paper,reqno]{amsart}
\usepackage{amsmath}
\usepackage{amssymb}
\usepackage{amsthm}
\usepackage{mathrsfs}
\usepackage{graphicx}
\usepackage{textcomp}
\newtheorem{thm}{Theorem}[section]
\newtheorem{prop}[thm]{Proposition}
\newtheorem{lemma}[thm]{Lemma}
\newtheorem{cor}[thm]{Corollary}
\newtheorem{assump}{Assumption}
\theoremstyle{definition}

\newtheorem{remark}[thm]{Remark}

\newtheorem*{prf}{Proof}

\numberwithin{equation}{section}

\def\dtworho#1{\frac{\partial^2#1}{\partial\rho^2}}

\def\PD#1#2{\partial_{#2}#1}

\def\bigo{\mathcal{O}}

\def\Ham{\mathcal{H}}
\def\h{\mathfrak{h}}
\def\dh{\delta\h}

\def\q{\mathfrak{q}}
\def\Q{\mathcal{Q}}

\def\tq{\tilde{\q}}
\def\tQ{\widetilde{\Q}}
\def\hq{\hat{\q}}
\def\hQ{\widehat{\Q}}
\def\eig#1#2{\lambda_{#1,#2}}
\def\eigone#1{\eig{1}{#1}}
\def\eigonepm#1{\eig{1}{#1,\pm}'}
\def\eigonep#1{\eig{1}{#1,+}'}
\def\eigonem#1{\eig{1}{#1,-}'}

\def\C{\mathcal{C}} % constant in the O(1)-term of the eigenvalue
\def\k{K} % The name of the constant
\def\dGdiff{\delta_0} % half the de Gennes derivative
\def\mom{\zeta} % the parameter name for the angular momentum
\def\hmom{\widehat{\mom}}
\def\tmom{\widetilde{\mom}}
\def\tzeromom{\tmom_0} % temp \mom_0
\def\tonemom{\tmom_1} % temp \mom_1
\def\para{\delta}

\def\hpsi{\widehat{\psi}}
\def\cpar{\varsigma}
\def\ccon{l}

\def\ed{\,\mathrm{d}}

\def\dG{\mathcal{G}}
\def\M{\mathcal{M}}
\def\Rreg{R_{\text{reg}}}
\def\tRreg{\widetilde{R}_{\text{reg}}}
\def\Rp{R^{+}}
\def\Rm{R^{-}}
\def\E{E_{0}}
\def\Ezero{\mathcal{E}_0}
\def\HH{\mathcal{H}}
\def\K{\mathcal{K}}
\def\Einf{E^N}
\def\Einfp{E^{N,+}}
\def\Einfm{E^{N,-}}
\def\Einfpm{E^{N,\pm}}

\def\Epm{E^{\pm}}
\def\Ep{E^{+}}
\def\tEpm{\widetilde{E}^{\pm}}
\let\phi=\varphi
\let\epsilon=\varepsilon
\DeclareMathOperator{\curl}{curl}
\DeclareMathOperator{\Div}{div}
\DeclareMathOperator{\dist}{dist}
\DeclareMathOperator{\dom}{Dom}
\DeclareMathOperator{\spec}{Spec}
\DeclareMathOperator{\supp}{supp}
\DeclareMathOperator{\Span}{span}
\DeclareMathOperator{\Cos}{Cos}

\title[Strong diamagnetism for the ball in three dimensions]%
{Strong diamagnetism for the ball\\ in three dimensions}

\author{S\o ren Fournais}
\author{Mikael Persson}
\address[S\o ren Fournais and Mikael Persson]{Aarhus University, Department of
  Mathematical Sciences, 1530 Ny Munkegade, 8000 Aarhus C, Denmark}
\email{fournais@imf.au.dk, mickep@imf.au.dk}

\begin{document}

\begin{abstract}
In this paper we give a detailed asymptotic formula for the lowest eigenvalue of
the magnetic Neumann Schr\"odinger operator in the ball in three dimensions with
constant magnetic field, as the strength of the magnetic field tends to
infinity. This asymptotic formula is used to prove that the eigenvalue is 
monotonically increasing for large values of the magnetic field.
\end{abstract}

\maketitle

\section{Introduction}
\subsection{The operator and main results}
Let $\Omega$ be the unit ball
\begin{equation*}
\Omega=\bigl\{x=(x_1,x_2,x_3)\in\mathbb{R}^3~\bigm|~|x|<1\big\}
\end{equation*} 
and let $\mathbf{B}$ be a constant magnetic field of
magnitude $B>0$ along the $x_3$ axis, with a corresponding choice of 
magnetic vector potential $\mathbf{A}$,
\begin{equation*}
\mathbf{B} = (0,0,B),\quad \mathbf{A}=\frac{B}{2}(-x_2,x_1,0).
\end{equation*}
We consider the magnetic Neumann Schr\"odinger operator
\begin{equation}\label{eq:neumannop}
\Ham(B)=(-i\nabla+\mathbf{A})^2
\end{equation}
with domain
\begin{equation}\label{eq:neumanncond}
\dom(\Ham(B))=\bigl\{\Psi\in W^{2,2}(\Omega)\mid
N(x)\cdot(-i\nabla+\mathbf{A})\Psi|_{\partial\Omega}=0\bigr\}.
\end{equation}
Here $N(x)$ is the interior unit normal to $\partial\Omega$. This
operator has compact resolvent and is semi-bounded from below, so it makes sense 
to enumerate its eigenvalues in an increasing order. For a self-adjoint operator 
$\Ham$ that is semi-bounded from below  we will use the notation $\eig{j}{\Ham}$ 
to denote its $j$th eigenvalue. In particular, we will write
\begin{equation*}
\eigone{\Ham(B)} = \inf\spec\bigl(\Ham(B)\bigr)
\end{equation*}
for the lowest eigenvalue of $\Ham(B)$. The first main theorem of this paper 
concerns the asymptotics of $\eigone{\Ham(B)}$ as $B\to\infty$.
\begin{thm}
\label{thm:main}
There exist constants $\lambda_j$, $j=0,\ldots,5$, and
$\hmom_0$, $\hmom_1$, $\hmom_2$, $\hmom_3$, $\dGdiff$, and $\C$
such that with
\begin{equation}\label{eq:momthree}
\mom_3(m,B) = m-\frac{B}{2}-\hmom_0\sqrt{B}-\hmom_1 B^{1/3}-\hmom_2 B^{1/6}.
\end{equation}
and
\begin{equation}\label{eq:deltab}
\Delta_B = \inf_{m\in\mathbb{Z}} \bigl|\mom_3(m,B)-\widehat{\mom}_3\bigr|
\end{equation}
it holds that
\begin{equation}
\label{eq:AsymptoticsMain}
\eigone{\Ham(B)}= B\sum_{j=0}^5 \lambda_j B^{-j/6}
+\dGdiff\Delta_B^2+\C+\bigo(B^{-1/6}),\quad \text{as } B\to\infty.
\end{equation}
\end{thm}

\begin{remark}
In the course of the proof we will obtain explicit expressions for the constants
in the theorem above, especially it holds that
\begin{equation}\label{eq:firstlambdas}
\lambda_0 = \Theta_0,\quad \lambda_1 = 0,\quad \text{and}\quad 
\lambda_2 = 2^{-2/3}\hat{\nu}_0\dGdiff^{1/3},
\end{equation}
which agrees with the asymptotics of $\eigone{\Ham(B)}$ that is given 
in~\cite{hemo} for more general domains $U\subset\mathbb{R}^3$ 
(see Theorem~\ref{thm:hemo}). The constants 
$\Theta_0$, $\hat{\nu}_0$ and $\dGdiff$ are well-known universal constants which
appear in the study of two model operators, see 
Appendix~\ref{sec:modeloperators}.
\end{remark}

Before stating the next main theorem it is worth noticing that the constants 
$\lambda_0=\Theta_0$ and $\dGdiff$ in the theorem satisfy $\frac12<\Theta_0<1$ 
and  $0<\dGdiff<1$, so especially, it holds that
\begin{equation}\label{eq:konstineq}
\Theta_0 - \frac12\dGdiff > 0.
\end{equation}

\begin{thm}
\label{thm:monotonicity}
Let $\dGdiff$ and $\lambda_0=\Theta_0$ be the constants from 
Theorem~\ref{thm:main}. The directional derivatives
\begin{equation*}
\eigonepm{\Ham(B)} = \lim_{\epsilon\to
0_{\pm}}\frac{\eigone{\Ham(B+\epsilon)}-\eigone{\Ham(B)}}{\epsilon}
\end{equation*}
exist and satisfy
\begin{align}
\label{eq:lplm}
\eigonep{\Ham(B)} 
& \leq \eigonem{\Ham(B)}, \quad \text{for all $B>0$},\\
\label{eq:lpineq}
\liminf_{B\to\infty} \eigonep{\Ham(B)} 
& \geq \Theta_0 - \frac{1}{2}\dGdiff>0,\quad \text{and}\\
\label{eq:lmineq}
\limsup_{B\to\infty} \eigonem{\Ham(B)} 
& \leq \Theta_0 + \frac{1}{2}\dGdiff.
\end{align}
In particular, the function $B\mapsto \eigone{\Ham(B)}$ is monotonically 
increasing for sufficiently large $B$.
\end{thm}

\subsection{Motivation}
\subsubsection{Strong diamagnetism}
Let $\Ham_U(B)$ denote the magnetic Neumann operator in a bounded and smooth 
domain $U\subset \mathbb{R}^n$, $n=2,3$. From the diamagnetic inequality 
(see~\cite{kato2}) it follows that
\begin{equation}
\eigone{\Ham_U(0)}\leq \eigone{\Ham_U(B)} 
\end{equation}
for all $B\geq 0$. One might ask if the stronger monotonicity
\begin{equation}
\label{eq:strongdia}
0<B_1<B_2\ \implies\ \eigone{\Ham_U(B_1)} \leq \eigone{\Ham_U(B_2)}
\end{equation}
holds. In~\cite{er1} counter-examples are given showing 
that~\eqref{eq:strongdia} does not hold in general \emph{for all} $B_1$ 
and $B_2$. The examples are given in $\mathbb{R}^2$, for constant magnetic 
field and the presence of a scalar potential, and for variable magnetic field 
without a scalar potential.

Lately the question whether there exist a $B_0$  such that 
$\eigone{\Ham_U(B)}$ is monotonically increasing for all $B>B_0$ has
been studied in detail. This is 
well-understood in two dimensions by now, with the final affirmative answer for 
regular domains in~\cite{fohe3,fohe5} and for domains with corners 
in~\cite{bofo}. We discuss below the progress in three dimensions so far, 
which motivates our analysis for the ball. 

To continue, we introduce some conditions on the domain $U$. Let 
$\Gamma\subset \partial U$ be
the set of all points on the boundary where the magnetic field $\mathbf{B}$ is
tangent to $\partial U$, i.e.
\begin{equation}\label{eq:Gamma}
\Gamma = \bigl\{x\in\partial U~\bigm|~\mathbf{B}\cdot N(x)=0\bigr\}.
\end{equation}

\begin{assump}
\label{assump:one}
Let $d$ denote the differential on $\partial U$. Then
\begin{equation*}
d(\mathbf{B}\cdot N(x))\neq 0,\quad \text{for all } x\in\Gamma.
\end{equation*}
\end{assump}
If this assumption holds then $\Gamma$ consists of a disjoint union of regular
curves. We can orient them and denote by $T(x)$ an oriented unit tangent vector
at $x\in\Gamma$. It is noted in~\cite{fohe6} that this implies that the magnetic
normal curvature $k_{n,B}(x)=K_x\bigl(T(x)\wedge
N(x),\frac{\mathbf{B}}{|\mathbf{B}|}\bigr)$ is non-zero on $\Gamma$. Here $K$ 
denotes the second fundamental form on $\partial U$.

\begin{assump}
\label{assump:two}
The set of points in $\Gamma$ where $\mathbf{B}$ is tangent to $\Gamma$ is
isolated.
\end{assump}

The following asymptotic formula of $\eigone{\Ham_U(B)}$ was proved 
in~\cite{hemo} (the upper bound was given in~\cite{pan}).

\begin{thm}\label{thm:hemo}
Let $U \subset\mathbb{R}^3$ be a bounded and smooth domain that satisfies 
Assumptions~\ref{assump:one} 
and~\ref{assump:two}. Then there exist constants $\Theta_0$, 
$\widehat{\gamma}_0>0$ and $\eta>0$ such that
\begin{equation}
\label{eq:hemo}
\eigone{\Ham_U(B)} = \Theta_0 B +\widehat{\gamma}_0 B^{2/3}+\bigo(B^{2/3-\eta}),
\quad \text{as } B\to\infty.
\end{equation}
\end{thm}

The constant $\Theta_0$ is the same as in Theorem~\ref{thm:main} and 
the constant $\widehat{\gamma}_0$ is given by
$\widehat{\gamma}_0=\inf_{x\in\Gamma}\widetilde{\gamma}_0(x)$ where
\begin{equation*}
\widetilde{\gamma}_0(x) = 2^{-2/3} \hat{\nu}_0 \dGdiff^{1/3}
|k_{n,B}(x)|^{2/3}\biggl(1+(\dGdiff-1)\Bigl|T(x)\cdot
\frac{\mathbf{B}}{|\mathbf{B}|}\Bigr|^2\biggr)^{1/3},
\end{equation*}
and $\dGdiff$ and $\hat{\nu}_0$ are fundamental constants given in
Appendix~\ref{sec:modeloperators}. We note that 
\begin{equation}\label{eq:hatgamma}
\widehat{\gamma}_0=2^{-2/3}\hat{\nu}_0\dGdiff^{1/3}
\end{equation}
when $U=\Omega$ is the unit 
ball which makes our Theorem~\ref{thm:main} compatible with 
Theorem~\ref{thm:hemo}.

Using the
expansion~\eqref{eq:hemo} of the lowest eigenvalue $\eigone{\Ham_U(B)}$, the
following monotonicity result was proved in~\cite{fohe6}.
\begin{thm}\label{thm:fohedia}
Let $U \subset\mathbb{R}^3$ be a bounded and smooth domain that satisfies 
Assumptions~\ref{assump:one}
and~\ref{assump:two}. Let $\{\Gamma_1,\ldots,\Gamma_n\}$ be the collection of
disjoint smooth curves making up $\Gamma$. Assume that for all $j$ there exists
$x\in\Gamma_j$ such that $\widetilde{\gamma}_0(x)>\widehat{\gamma}_0$. Then the 
function $B\mapsto \eigone{\Ham_U(B)}$ is strictly increasing for sufficiently 
large $B$.
\end{thm}

This shows that~\eqref{eq:strongdia} holds for large values of $B_1$ and $B_2$.
Even though the Assumptions~\ref{assump:one} and~\ref{assump:two} are fulfilled
for the ball, the assumption on $\widetilde{\gamma}_0$ in 
Theorem~\ref{thm:fohedia} is not. Indeed, for the unit ball $\Omega$, the
set $\Gamma$ consists of the equator $\{x\in\partial\Omega\mid x_3=0\}$ 
and the function $\widetilde{\gamma}_0(x)$ is
constant
\begin{equation*}
\widetilde{\gamma}_0(x)\equiv 2^{-2/3} \hat{\nu}_0 \dGdiff^{1/3}
,\quad x\in\Gamma.
\end{equation*}

\subsubsection{Superconductivity} We consider superconductivity in the
Ginzburg-Landau model. For a superconducting material of shape
$U$ subject to an external magnetic field $\kappa \sigma \beta$,
with $\beta = (0,0,1)$, the Ginzburg-Landau energy functional is as follows,
\begin{multline}
  \label{eq:1}
  {\mathcal E}_{\kappa,\sigma}(\Psi, \mathbf{a}) =
  \int_{U}|(-i\nabla +\kappa \sigma \mathbf{a})\Psi|^2 - \kappa^2 |\Psi|^2
  +\frac{\kappa^2}{2} |\Psi|^4 \ed x \\+
(\kappa \sigma)^2 \int_{{\mathbb R}^3} |\curl \mathbf{a} - \beta|^2 \ed x.
\end{multline}
Here $\kappa>0$ is a material dependent parameter called the
Ginzburg-Landau parameter. For given $\kappa$, the parameter $\sigma$
measures the strength of the external magnetic field. The
function $\Psi\in H^1(U)$ is in this context called an order parameter and 
$|\Psi|$ measures
the local superconducting properties (density of Cooper pairs) of the
material. Finally $\mathbf{a} \in H^1_{\rm loc}({\mathbb R}^3)$ is the
induced magnetic vector potential. In order to get a well-defined
minimization problem, one uses gauge invariance to restrict to vector
potentials satisfying $\Div \mathbf{a} = 0$, and impose the finite energy
condition (see \cite{giti} for details),
\begin{align*}
  \int_{{\mathbb R}^3} |\curl \mathbf{a} - \beta|^2\ed x < \infty.
\end{align*}
A state $(\Psi, \mathbf{a})$ where $\Psi \equiv 0$ and $\curl \mathbf{a} =
\beta$ is called trivial.

The analysis of magnetic ground state eigenvalues described above is
relevant in superconductivity for the understanding of the loss of
superconductivity in the presence of strong magnetic fields. The value
of the
magnetic field strength at which the material loses its
superconducting properties, i.e., the minimizers of the Ginzburg-Landau
functional have $\Psi \equiv 0$, is called \emph{the third critical field}. 
The calculation of this critical field has a long history,
see \cite{giph,best,lupa1,lupa4,hepa,fohe3,fohe5,fohebook}.

In \cite{fohe6,fohebook}
it was proved that, for $\kappa$ sufficiently
large, the following sets are equal
\begin{align}
  \label{eq:2}
  {\mathcal N}(\kappa) &:= \bigl\{ \sigma >0~\bigm|~{\mathcal E}_{\kappa,
    \sigma} \text{ has a non-trivial minimizer}\bigr\} \nonumber \\
&\hphantom{:}=\bigl\{\sigma >0~\bigm|~{\mathcal E}_{\kappa,
    \sigma} \text{ has a non-trivial stationary point}\bigr\}  \\
&\hphantom{:}=\bigl\{\sigma >0~\bigm|~\lambda_{1,{\mathcal H}_{U}(\kappa
  \sigma)} < \kappa^2 \bigr\}.\nonumber
\end{align}
In the case where $U$ is the unit ball $\Omega$, we can use the
monotonicity result of Theorem~\ref{thm:monotonicity} to conclude
that, for $\kappa$ sufficiently large,
\begin{align}
  \label{eq:3}
  \bigl\{\sigma >0~\bigm|~\eigone{\Ham_{\Omega}(\kappa
  \sigma)} < \kappa^2 \bigr\} = (0, H_{C_3}(\kappa)).
\end{align}
Here $\sigma = H_{C_3}(\kappa)$ is the unique solution to the
equation
\begin{equation*}
  \eigone{\Ham_{\Omega}(\kappa\sigma)} = \kappa^2.
\end{equation*}
Hereby, we get a complete determination of the third critical field,
$H_{C_3}(\kappa),$ for large values of $\kappa$.
Upon inserting the asymptotics \eqref{eq:AsymptoticsMain}, one gets a
six-term asymptotic expansion of $ H_{C_3}(\kappa)$ for the ball,
\begin{multline*}
H_{C_3}(\kappa)=\frac{\kappa}{\Theta_0}
-\frac{\widehat{\gamma}_0}{\Theta_0^{5/3}}\kappa^{1/3}
-\frac{\lambda_3}{\Theta_0^{3/2}}
+\biggl(\frac{2\widehat{\gamma}_0}{3\Theta_0^{7/3}}
-\frac{\lambda_4}{\Theta_0^{4/3}}\biggr)\kappa^{-1/3}
+\biggl(\frac{7\lambda_3\widehat{\gamma}_0}{6\Theta_0^{13/6}}
-\frac{\lambda_5}{\Theta_0^{7/6}}\biggr)\kappa^{-2/3}\\
+\biggl(\frac{\lambda_3^2}{2\Theta_0^2}
+\frac{\lambda_4\widehat{\gamma}_0}{\Theta_0^2}
-\frac{\widehat{\gamma}_0^3}{3\Theta_0^3}
-\frac{\Delta_{\kappa H_{C_3}(\kappa)}^2+\C}{\Theta_0}\biggr)\kappa^{-1}
+\bigo\bigl(\kappa^{-4/3}\bigr)
\end{multline*}
Thus, ${\mathcal N}(\kappa)$ is an interval (for large $\kappa$) both
in the case where
$U$ satisfies the assumptions of Theorem~\ref{thm:fohedia} and in the
case where $U$
is a ball. It remains an interesting open question to prove this
result in general, i.e., to prove strong diamagnetism for general (smooth)
domains in ${\mathbb R}^3$.

\subsection{Organization of the paper}

The main part of this paper is devoted to the proof of Theorem~\ref{thm:main}. 
The proof is divided into several parts: 

We denote by $\Ham_m(B)$ the operator 
$\Ham(B)$ restricted to angular momentum $m\in\mathbb{Z}$. 
In Section~\ref{sec:lower} we show
a lower bound for $\Ham_m(B)$. First we show in Lemmas~\ref{lem:firstmombound} 
and~\ref{lem:bhalf} that (in terms of $m$ and $B$) 
either the first eigenvalues of $\Ham_m(B)$ are too large to be compatible
with~\eqref{eq:AsymptoticsMain}, or we can reduce 
$\Ham_m(B)$ to an effective operator $\Q_m(B)$, satisfying, as $B\to\infty$,
\begin{equation}\label{eq:helpupperone}
\eig{j}{\Ham_m(B)} = B\eig{j}{\Q_m(B)} + \bigo(B^{1/2}),\quad \text{for }j=1,2.
\end{equation}
For the values of $m$ and $B$ such that~\eqref{eq:helpupperone} holds, we prove
a lower bound for $\Q_m(B)$ in Proposition~\ref{prop:rholoc}.

In Section~\ref{sec:newloc} we use the lower bound from Section~\ref{sec:lower} 
to give localization properties of eigenfunctions of $Q_m(B)$. These are
used in Section~\ref{sec:gap} to obtain a spectral gap formula for $\Q_m(B)$, 
which together with~\eqref{eq:helpupperone} implies a spectral gap formula for 
$\Ham_m(B)$, showing that for some $\gamma>0$ it holds that 
(still under some restrictions on $m$ and $B$)
\begin{equation}\label{eq:SpecGap}
\eig{2}{\Ham_m(B)}\geq
\Theta_0 B+
(\widehat{\gamma}_0+\gamma) B^{2/3} +\bigo(B^{7/12})
,\quad\text{as }B\to\infty.
\end{equation}
In Section~\ref{sec:upper}, Theorem~\ref{thm:goodtrial}(iii) 
we use the Gru{\v{s}}in method to, for certain $m$ 
and $B$, calculate a trial state that together with the spectral gap formula
gives upper and lower bounds on $\eigone{\Ham_m(B)}$ 
which are compatible with~\eqref{eq:AsymptoticsMain}. We also give two 
alternative trial states, in Theorem~\ref{thm:goodtrial}(i) and~(ii), 
for values of $m$ that are further away from the optimal choice.

In Section~\ref{sec:newlower} we show that $\eigone{\Ham_m(B)}$ is larger for 
the values of $m$ and $B$ not treated in Theorem~\ref{thm:goodtrial}(iii). 
Depending on $m$ and $B$ we use different methods to achieve this. For $m$ and 
$B$ which are far from the optimal region we use (a refined version of) the 
lower bound from Section~\ref{sec:lower}. For $m$ and $B$ that are closer to 
the optimal region we use the trial state from 
Theorem~\ref{thm:goodtrial}(i) and (ii) which by the spectral gap 
formula~\eqref{eq:SpecGap} must be $\eigone{\Ham_m(B)}$, but which is strictly 
greater than the ones obtained in Theorem~\ref{thm:goodtrial}(iii).
Finally, in Section~\ref{sec:proof} we minimize the eigenvalue found in 
Theorem~\ref{thm:goodtrial}(iii), which proves Theorem~\ref{thm:main}.

The proof of Theorem~\ref{thm:monotonicity} is a consequence of 
Theorem~\ref{thm:main}, using a perturbation argument from~\cite{fohebook}. 
The details are given in Section~\ref{sec:monotonicity}. 

We start by introducing the new coordinates and some quadratic forms and 
operators.

\subsection{New coordinates, auxiliary operators and quadratic forms}

We conclude this first section by introducing the coordinates we will work in 
and the different quadratic forms we will work with. First, we switch to 
spherical coordinates $(x_1,x_2,x_3)\mapsto
(r,\phi,\theta)$,
\begin{equation*}
\left\{
\begin{aligned}
x_1 &= r \cos\phi \sin\theta,\\
x_2 &= r \sin\phi \sin\theta,\\
x_3 &= r \cos\theta,
\end{aligned}\right.\qquad 0<r\leq1,\ 0<\theta<\pi, 0\leq \phi < 2\pi.
\end{equation*}
We decompose the Hilbert space as
\begin{align*}
L^2(\Omega)&\cong 
L^2\big((0,1)\times (0,\pi),r^2 \sin\theta \ed r \ed \theta\big) \otimes
L^2(\mathbb{S}^1,d\phi)\\
&\cong \bigoplus_{m=-\infty}^\infty L^2\big((0,1)\times (0,\pi),r^2 \sin\theta 
\ed r \ed \theta\big) \otimes \frac{e^{-im\phi}}{\sqrt{2\pi}},
\end{align*}
that is, for a function $\Psi\in L^2(\Omega)$, we write
\begin{equation*}
\Psi(r,\phi,\theta)=\sum_{m\in\mathbb{Z}} \psi_m(r,\theta)
\frac{e^{-im\phi}}{\sqrt{2\pi}},
\end{equation*}
where $\psi_m\in L^2\big((0,1)\times (0,\pi),r^2 \sin\theta 
\ed r \ed \theta\big)$. Next, we write the operator $\Ham(B)$ corresponding to 
this decomposition as
\begin{equation*}
\Ham(B) = \bigoplus_{m=-\infty}^\infty \Ham_m(B)\otimes 1
\end{equation*}
where $\Ham_m(B)$ is the self-adjoint operator acting in $L^2\big((0,1)\times
(0,\pi),r^2\sin\theta \ed r \ed \theta\big)$, given by
\begin{equation*}
\Ham_m(B)
=-\PD{}{r}^2-\frac{2}{r}\PD{}{r}-
\frac{1}{r^2}\PD{}{\theta}^2 -
\frac{1}{r^2\tan\theta}\PD{}{\theta}+\Bigl(\frac{Br\sin\theta}{2} - 
\frac{m}{r\sin\theta}\Bigr)^2,
\end{equation*}
with Neumann boundary condition at $r=1$.
In the continuation we skip the subscript $m$ on $\psi_m$ and write just $\psi$.
Inspired by~\cite{hemo} we introduce the new scaled coordinates $(\tau,\rho)$ as
\begin{equation}
\label{eq:newcoord}
\left\{%
\begin{aligned}
\tau &=\sqrt{B}(1-r),\\
\rho &= \sqrt[3]{B}(\theta-\pi/2),
\end{aligned}
\right.
\end{equation}
with corresponding new domain
\begin{equation*}
\Omega(B)=\Bigl\{(\tau,\rho)\mid 0<\tau<\sqrt{B},\
-\frac{\pi}{2}\sqrt[3]{B}<\rho<\frac{\pi}{2}\sqrt[3]{B}\Bigr\}. 
\end{equation*}
In fact, for a point $x$ in $\Omega$, $\tau=\sqrt{B}\dist(x,\partial\Omega)$ is
equal to the (scaled) distance to the boundary and for a point 
$x\in\partial\Omega$ we have 
$\rho=\sqrt[3]{B}\dist_{\partial{\Omega}}(x,\Gamma)$, the (scaled) distance
along the boundary to the equator.

The quadratic form corresponding to $\frac{1}{B}\Ham_m$ (the prefactor $1/B$ is 
just for convenience) transforms into the 
quadratic form
\begin{align}
 \tq_m[\psi] &= \int_{\Omega(B)} \biggl[\left|\PD{\psi}{\tau}\right|^2 +
\frac{B^{-1/3}}{(1-B^{-1/2}\tau)^2}\left|\PD{\psi}{\rho}\right|^2 
\label{eq:qtilde}\\
&\quad + \frac{1}{B}\left(\frac{B (1-B^{-1/2}\tau)\cos(B^{-1/3}\rho)}{2} -
\frac{m}{(1-B^{-1/2}\tau)\cos(B^{-1/3}\rho)}\right)^2|\psi|^2\biggr]\times
\nonumber\\
&\qquad \times (1-B^{-1/2}\tau)^2 \cos(B^{-1/3}\rho) B^{-5/6}\ed\tau \ed\rho
\nonumber
\end{align}
in the Hilbert space
\begin{equation*}
L^2\Bigl(\Omega(B),(1-B^{-1/2}\tau)^2 \cos(B^{-1/3}\rho) B^{-5/6}\ed\tau
\ed\rho\Bigr).
\end{equation*}

We apply the unitary transform $\mathcal{U}\psi=B^{-5/12}\psi$ to get rid of
the factor $B^{-5/6}$ in the measure and work in the Hilbert space
\begin{equation*}
L^2\Bigl(\Omega(B),(1-B^{-1/2}\tau)^2 \cos(B^{-1/3}\rho) \ed\tau\ed\rho\Bigr)
\end{equation*}
instead. We will, by abuse of notation, continue to write $\psi$ instead 
of $\mathcal{U}\psi$. We denote 
by $\tQ_m(B)$ the operator corresponding to the quadratic form $\tq_m$.

Next we want to define a quadratic form $\hq_m$ by the same integral expression
as for $\tq_m$ but in the Hilbert space
$L^2(\mathbb{R}^2_+,\ed\tau \ed\rho)$, where 
\begin{equation*}
\mathbb{R}^2_+=\{(\tau,\rho)\mid 0<\tau<\infty,\ -\infty<\rho<\infty\}.
\end{equation*}
However, since neither $1-B^{-1/2}\tau$ nor $\cos(B^{-1/3}\rho)$ are strictly 
positive in $\mathbb{R}^2_+$, we make a technical modification to be able to
talk about the corresponding operator. We define
smooth functions $\ell:\mathbb{R}_+\to\mathbb{R}_+$ and 
$\Cos:\mathbb{R}\to\mathbb{R}$ that satisfy
\begin{equation}\label{eq:ellcos}
\ell(x)=
\begin{cases}
x, & x\leq\frac13,\\
\frac12, & x\geq\frac12,
\end{cases}
\quad \text{and}\quad
\Cos(x)=
\begin{cases}
\cos(x), & |x|\leq\frac\pi4,\\
\frac12, & |x|\geq\frac\pi3,
\end{cases}
\end{equation}
and such that $\ell$ is monotonically increasing in the interval 
$(\frac13,\frac12)$ and $\Cos$ is even and monotonically decreasing in 
the interval $(\frac\pi4,\frac\pi3)$. 
The quadratic form $\hq_m$ is defined by
\begin{multline}
\label{eq:qhat}
\hq_m[\psi] = \int_{\mathbb{R}^2_+} \biggl[\left|\PD{\psi}{\tau}\right|^2 +
\frac{B^{-1/3}}{\bigl(1-\ell(B^{-1/2}\tau)\bigr)^2}
\left|\PD{\psi}{\rho}\right|^2 \\
+ \frac{1}{B}
\biggl(\frac{B \bigl(1-\ell(B^{-1/2}\tau)\bigr)\Cos(B^{-1/3}\rho)}{2} -
\frac{m}{\bigl(1-\ell(B^{-1/2}\tau)\bigr)\Cos(B^{-1/3}\rho)}\biggr)^2
|\psi|^2\biggr]\\
\times \bigl(1-\ell(B^{-1/2}\tau)\bigr)^2 \Cos(B^{-1/3}\rho) \ed\tau \ed\rho.
\end{multline}
We denote by $\hQ_m(B)$ the self-adjoint operator that corresponds to $\hq_m$. 
It is an operator in 
$L^2\Bigl(\mathbb{R}^2_+,
\bigl(1-\ell(B^{-1/2}\tau)\bigr)^2 \Cos(B^{-1/3}\rho)\Bigr)$ 
with Neumann condition at $\tau=0$. An integration by parts, show that it acts 
as
\begin{multline}\label{eq:hQm}
\hQ_m(B)= 
-\PD{}{\tau}^2
+\frac{2\ell'(B^{-1/2}\tau)B^{-1/2}}{(1-\ell(B^{-1/2}\tau))}\PD{}{\tau}\\
-\frac{B^{-1/3}}{\bigl(1-\ell(B^{-1/2}\tau)\bigr)^2}
\Bigl(\PD{}{\rho}^2
+\frac{\Cos'(B^{-1/3}\rho)B^{-1/3}}{\Cos(B^{-1/3}\rho)}\PD{}{\rho}\Bigr)\\
+\frac{1}{B}
\biggl(\frac{B \bigl(1-\ell(B^{-1/2}\tau)\bigr)\Cos(B^{-1/3}\rho)}{2} 
-\frac{m}{\bigl(1-\ell(B^{-1/2}\tau)\bigr)\Cos(B^{-1/3}\rho)}\biggr)^2
\end{multline}

Finally, we also define the quadratic form $\q_m$ in $L^2(\mathbb{R}^2_+,\ed\tau
\ed\rho)$ by
\begin{equation}
\label{eq:qm}
\q_m[\psi] = \int_{\mathbb{R}^2_+} \left|\PD{\psi}{\tau}\right|^2+\left(\tau +
\frac{1}{\sqrt{B}}(m-B/2) + B^{-1/6}\frac{\rho^2}{2}\right)^2|\psi|^2 +
B^{-1/3}\left|\PD{\psi}{\rho}\right|^2\ed\tau \ed \rho
\end{equation}
with corresponding self-adjoint operator $\Q_m(B)$ with Neumann boundary 
condition for $\tau=0$.

%
%
%
%
%
%            LOWER        BOUNDS
%
%
%
%
%
\section{A rough lower bound}
\label{sec:lower}
In this section we recall the localization formulas of the lowest eigenfunction 
of $\Ham(B)$ obtained in~\cite{hemo} and written out in detail in~\cite{fohe5}, 
use them to reduce the study of $\Ham_m(B)$ to the study of $\Q_m(B)$, and 
give a rough lower bound of its quadratic form $\q_m$ in 
Proposition~\ref{prop:rholoc}.

\begin{thm}\label{thm:helpagmon}
Suppose that $U\subset\mathbb{R}^3$ satisfies the
Assumptions~\ref{assump:one} and~\ref{assump:two} and that $\Psi$ satisfies
$\Ham(B)\Psi=\lambda\Psi$ with 
$\lambda(B)\leq \Theta_0B+\omega B^{2/3}$. 
Then there exist positive constants $a_1$, $a_2$,
$d_0$, $C$ and $B_0$ such that
\begin{equation*}
\int_{U} e^{2a_1 B^{1/2}\dist(x,\partial\Omega)}\bigl(|\Psi|^2 +
B^{-1}|(-i\nabla+\mathbf{A})\Psi|^2\bigr) \ed x \leq C\|\Psi\|^2
\end{equation*}
and
\begin{multline*}
\int_{\{\dist(x,\partial U)\leq d_0\}} e^{2a_2
B^{1/2}\dist_{\partial U}(x,\Gamma)^{3/2}}\left(|\Psi|^2 +
B^{-1}|(-i\nabla+\mathbf{A})\Psi|^2\right) \ed x \\
\leq C e^{CB^{1/8}}\|\Psi\|^2
\end{multline*}
for all $B\geq B_0$.
\end{thm}

From now on we assume that $U=\Omega$ is the unit ball. Then the set $\Gamma$, 
introduced in~\eqref{eq:Gamma}, consists of
the equator and we can extend the distance function
$\dist_{\partial\Omega}(x,\Gamma)$ to all of $\Omega$ (except the origin) as
$\dist_{\partial\Omega}(x,\Gamma)=\dist_{\partial\Omega}(\hat{x},\Gamma)$ where
$\hat{x}=\frac{x}{|x|}\in\partial\Omega$. By the
exponential decay away from the boundary, the second inequality in
Theorem~\ref{thm:helpagmon} is then valid with the integral on the left-hand
side being over all of $\Omega$, with possible changes of
the constants. We will use the following corollary.
\begin{cor}
\label{cor:agmon}
Suppose that $\Psi$ satisfies $\Ham(B)\Psi=\lambda\Psi$ with
$\lambda(B)\leq \Theta_0B+\omega B^{2/3}$. Then for all
$n\in\mathbb{N}$ there exist positive constants $C_n$ and $B_n$ such that
\begin{equation}
\label{eq:distboundary}
\int_{\Omega} \dist(x,\partial\Omega)^n
\left(|\Psi|^2+B^{-1}|(-i\nabla+\mathbf{A})\Psi|^2\right) \ed x \leq
C_nB^{-n/2}\|\Psi\|^2
\end{equation}
and
\begin{equation}
\label{eq:distequator}
\int_{\Omega} \dist_{\partial\Omega}(x,\Gamma)^{n}\left(|\Psi|^2 +
B^{-1}|(-i\nabla+\mathbf{A})\Psi|^2\right) \ed x \leq C_n B^{-n/4}\|\Psi\|^2
\end{equation}
for all $B\geq B_n$.
\end{cor}

\begin{remark}
The order in~\eqref{eq:distequator} is not optimal. The calculations below 
indicate that the same estimate is true with $B^{-n/3}$ instead of $B^{-n/4}$ 
in the right-hand-side.
\end{remark}

\noindent Let $0<\epsilon<1/12$ be given. We introduce a smooth cut-off function
$0\leq \chi_B \leq 1$ such that
\begin{equation}
\label{eq:cutoff}
\chi_B =
\begin{cases}
1, & \text{if $\dist(x,\partial\Omega)\leq B^{-1/2+\epsilon}$ and 
$\dist_{\partial\Omega}(x,\Gamma) \leq B^{-1/4+\epsilon}$},\\
0, & \text{if $\dist(x,\partial\Omega)\geq 2B^{-1/2+\epsilon}$ or 
$\dist_{\partial\Omega}(x,\Gamma) \geq 2B^{-1/4+\epsilon}$},
\end{cases}
\end{equation}
and such that $|\nabla\chi_B|\leq CB^{1/2-\epsilon}$ for some $C>0$.

\begin{lemma}
\label{lem:cutoff}
Suppose that $\Psi$ satisfies $\Ham(B)\Psi=\lambda\Psi$ with
$\lambda(B)\leq \Theta_0B+\omega B^{2/3}$. For any $N>0$ it
holds that
\begin{displaymath}
\int_\Omega |(-i\nabla +\mathbf{A})\Psi|^2 \ed x = \int_\Omega |(-i\nabla
+\mathbf{A})(\chi_B\Psi)|^2 \ed x + 
\bigo(B^{-N})\|\Psi\|^2,\quad\text{as}\quad B\to\infty.
\end{displaymath}
\end{lemma}

\begin{prf}
This follows by commuting $(-i\nabla-\mathbf{A})$ and $\chi_B$ and using
Corollary~\ref{cor:agmon}.\qed
\end{prf}

We remind the reader of the quadratic forms $\tq_m$ and $\hq_m$, introduced 
in~\eqref{eq:qtilde} and~\eqref{eq:qhat}, with corresponding self-adjoint 
operators $\tQ_m(B)$ and $\hQ_m(B)$. First we note that
\begin{equation}
\inf \spec\bigl(\Ham(B)\bigr) 
= \inf_{m\in\mathbb{Z}} \inf \spec\bigl(\Ham_m(B)\bigr) 
=B\inf_{m\in\mathbb{Z}} \inf \spec\bigl(\tQ_m(B)\bigr).
\end{equation}

We use Lemma~\ref{lem:cutoff} to reduce the study of $\Ham(B)$ to the study of 
the quadratic form $\hq_m$ in the half-space $\mathbb{R}_+^2$. 
We will denote by $\psi_B$ the function 
\begin{equation*}
\psi_B=\chi_B\psi,
\end{equation*}
where $\chi_B$ is the cut-off function from~\eqref{eq:cutoff}. Notice that 
$\psi$, and by consequence $\psi_B$, depends on $m$, but we do not include this
in the notation.
\begin{lemma}\label{lem:psicutoff}
Assume that $\psi_1$ and $\psi_2$ satisfies $\tQ_m(B)\psi_j=\lambda_j(B)\psi_j$ 
with $\lambda_j(B)\leq \Theta_0B+\omega B^{2/3}$. For any number $N>0$ there
exist constants $B_N$ and $C_N$ (independent of $m$) such that if $B>B_N$ then
\begin{equation*}
\bigl|\tq_m(\psi_1,\psi_2)-\hq_m(\chi_B\psi_1,\chi_B\psi_2)\bigr| 
\leq C_N B^{-N}\|\psi_1\|\cdot\|\psi_2\|.
\end{equation*}
\end{lemma}

\begin{prf}
This follows from Lemma~\ref{lem:cutoff}, transforming to the coordinates
$(\tau,\rho)$.\qed
\end{prf}

We recall that $0<\epsilon<1/12$ and note that 
\begin{equation}\label{eq:supppsib}
\supp \psi_B \subset \bigl\{ (\tau,\rho) \mid 0<\tau<2B^{\epsilon},\ 
-2B^{1/12+\epsilon}<\rho<2B^{1/12+\epsilon}\bigr\},
\end{equation}
which  implies
that $B^{-1/2}\tau\leq 2B^{\epsilon-1/2}$ and $|B^{-1/3}\rho|\leq
2B^{\epsilon-1/4}$ on the support of $\psi_B$. 

\begin{lemma}\label{lem:taurhoest}
Assume that $\psi$ satisfies $\tQ_m(B)\psi=\lambda(B)\psi$ with
$\lambda(B)\leq \Theta_0B+\omega B^{2/3}$. For any number $n>0$
there exist constants $B_n$ and $C_n$ (independent of $m$) such that if $B>B_n$ 
then
\begin{equation}
\label{eq:tauest}
\int_{\mathbb{R}^2_+} \tau^n \bigl(|\psi_B|^2+|\PD{\psi_B}{\tau}|^2 +
B^{-1/3}|\PD{\psi_B}{\rho}|^2 \bigr)\ed\tau \ed\rho \leq C_n \|\psi_B\|^2
\end{equation}
and
\begin{equation}
\label{eq:rhoest}
\int_{\mathbb{R}^2_+} |\rho|^{n}\bigl(|\psi_B|^2+|\PD{\psi_B}{\tau}|^2 +
B^{-1/3}|\PD{\psi_B}{\rho}|^2 \bigr)\ed\tau \ed\rho 
\leq C_n B^{n/12}\|\psi_B\|^2.
\end{equation}
\end{lemma}
\begin{prf}
This follows directly from Corollary~\ref{cor:agmon}, by transforming into the
new coordinates $(\tau,\rho)$.
\qed
\end{prf}
We will use the estimates in Lemma~\ref{lem:taurhoest} to reduce the values of
the angular momentum $m$ that we must consider, which in the end will enable us
to study the effective quadratic form $\q_m$ instead of $\hq_m$.  Let 
$\tzeromom=\frac{1}{\sqrt{B}}(m-B/2)$. 
\begin{lemma}\label{lem:firstmombound}
Assume that $\psi$ satisfies $\tQ_m(B)\psi=\lambda(B)\psi$ with
$\lambda(B)\leq \Theta_0B+\omega B^{2/3}$. There exist positive
constants $\widetilde{D}$ and $B_0$ such that if $|\tzeromom|>\widetilde{D}$ and
$B>B_0$ then 
\begin{equation}
\label{eq:bigenergy}
\hq_m[\psi_B] \geq \frac{1}{2}\tzeromom^2\|\psi_B\|^2.
\end{equation}
\end{lemma}

\begin{prf}
We expand the potential in $\hq_m$ and collect the terms in front of the
different degrees of $\tzeromom$. By~\eqref{eq:ellcos} and ~\eqref{eq:supppsib}
we see that if 
$B>\max\bigl((8/\pi)^{4/(1-4\epsilon)},6^{(2/(1-2\epsilon))}\bigr)$ then it 
holds that $\ell(B^{-1/2}\tau)=B^{-1/2}\tau$ and 
$\Cos(B^{-1/3}\rho)=\cos(B^{-1/3}\rho)$ on the support of $\psi_B$, and so
we can write the potential in $\hq_m$ as
\begin{equation*}
\biggl[\tzeromom-
\biggl(\frac{\sqrt{B}(1-B^{-1/2}\tau)^2\cos^2(B^{-1/3}\rho)}{2}
-\frac{\sqrt{B}}{2}\biggr)\biggr]^2\frac{1}{\cos(B^{-1/3}\rho)}
\end{equation*}
Using that $0<\cos(B^{-1/3}\rho)<1$ on the support of $\psi_B$ and the general
inequality $(x-y)^2\geq \frac34x^2-\frac13y^2$, valid for real $x$ and $y$, we
find that $\hq_m[\psi_B]$ is bounded from below by
\begin{equation*}
\int_{\mathbb{R}^2_+}
\biggl[\frac{3}{4}\tzeromom^2-
\frac{1}{3}\biggl(\frac{\sqrt{B}(1-B^{-1/2}\tau)^2\cos^2(B^{-1/3}\rho)}{2}
-\frac{\sqrt{B}}{2}\biggr)^2\frac{1}{\cos(B^{-1/3}\rho)}\biggr]
|\psi_B|^2 \ed\tau\ed\rho.
\end{equation*}
Using~\eqref{eq:tauest} and~\eqref{eq:rhoest} we get the existence of a
constant $C$ such that
\begin{equation*}
\int_{\mathbb{R}^2_+}
\biggl(\frac{\sqrt{B}(1-B^{-1/2}\tau)^2\cos^2(B^{-1/3}\rho)}{2}
-\frac{\sqrt{B}}{2}\biggr)^2\frac{1}{\cos(B^{-1/3}\rho)}
|\psi_B|^2 \ed\tau\ed\rho \leq C \|\psi_B\|^2.
\end{equation*}
This clearly implies~\eqref{eq:bigenergy}.\qed
\end{prf}

By Lemma~\ref{lem:firstmombound} above we need only to consider bounded 
$\tzeromom$. This enables to study the quadratic form $\q_m$ instead of
$\hq_m$. We assume that $|\tzeromom|<\widetilde{D}_0$ for some 
$\widetilde{D}_0>0$ and also let $D_0=\widetilde{D}_0+|\hmom_0|$ so that the
inequality $|\tzeromom-\hmom_0|<D_0$ holds. Here $\hmom_0=\xi_0$ is the constant
from Lemma~\ref{lem:thetazero}.

\begin{lemma}\label{lem:bhalf}
Assume that $\psi_1$ and $\psi_2$ satisfies $\tQ_m(B)\psi_j=\lambda(B)\psi_j$ 
with $\lambda_j(B)\leq \Theta_0B+\omega B^{2/3}$. If 
$|\tzeromom|<\widetilde{D}_0$ for some constant $\widetilde{D}_0>0$ then there 
exist constants $C>0$ and $B_0>0$ (independent of $m$ and $B$) such that
\begin{equation*}
\bigl|\hq_m(\chi_B\psi_1,\chi_B\psi_2) - \q_m(\chi_B\psi_1,\chi_B\psi_2)\bigr|
\leq C B^{-1/2}\|\psi_1\|\cdot\|\psi_2\|
\end{equation*}
for $B>B_0$.
\end{lemma}

\begin{prf}
This follows by expanding the terms in $\hq_m$ and estimating
using~\eqref{eq:tauest} and~\eqref{eq:rhoest}.\qed
\end{prf}

Next, we introduce 
\begin{equation}\label{eq:tonemom}
\tonemom=(\tzeromom-\hmom_0)B^{1/6}
=\Bigl(\frac{1}{\sqrt{B}}(m-B/2)-\hmom_0\Bigr)B^{1/6},
\end{equation}
and note that if $|\tzeromom|<\widetilde{D}_0$ then
\begin{equation}
\label{eq:bdddelta}
|\tonemom|\leq D_0 B^{1/6}.
\end{equation}
With this notation, the form $\q_m$ reads
\begin{equation*}
\q_m[\psi] = \int_{\mathbb{R}^2_+} \bigl|\PD{\psi}{\tau}\bigr|^2 +
\biggl(\tau+\hmom_0+B^{-1/6}\Bigl(\tonemom+\frac{\rho^2}{2}\Bigr)\biggr)^2
|\psi|^2 + B^{-1/3}\bigl|\PD{\psi}{\rho}\bigr|^2 \ed\tau \ed\rho.
\end{equation*}

Given constants $C_1$, $C_2$ and $M$ we define the function $W_B$ as
\begin{equation}\label{eq:WB}
W_B(\rho)=
\begin{cases}
\Theta_0 + C_1 B^{-1/3}, & |\rho|<M,\\
\Theta_0 + C_2 M^4 B^{-1/3}, & |\rho|\geq M.
\end{cases}
\end{equation}

\begin{prop}
\label{prop:rholoc}
Assume that $|\tonemom|\leq D_0 B^{1/6}$. 
There exist positive constants $C_1$, $C_2$, $\widetilde{M}$ and $B_0$ such that
for all $m\in\mathbb{Z}$
\begin{equation*}
\q_m[\psi] \geq \int_{\mathbb{R}^2_+} W_B(\rho) |\psi|^2 \ed\tau \ed\rho
\end{equation*}
for all $B>B_0$ and $\psi\in\dom(\q_m)$ if $M>\widetilde{M}$.
\end{prop}

Before proving this proposition we emphasize that these estimates do not only 
hold for ground states, but for all functions $\psi$ in the domain of $\q_m$.

\begin{prf}
We will prove a slightly stronger statement than the one in 
Proposition~\ref{prop:rholoc}. Let $D>0$ be a large number, to be specified 
below. In the following lemmas we consider the three cases:
\begin{itemize}
\item[(1)] $\tonemom>D$ as $B>B_1$ for some $B_1>0$ (Lemma~\ref{lem:rholocone}),
\item[(2)] $\tonemom<-D$ as $B>B_2$ for some $B_2>0$ 
(Lemma~\ref{lem:rholoctwo}),
\item[(3)] $|\tonemom|\leq D$ for all $B$ (Lemma~\ref{lem:rholocthree}).
\end{itemize}
In fact, for the cases $(1)$ and $(2)$ we will prove stronger estimates.
\qed
\end{prf}

\begin{lemma}
\label{lem:rholocone}
Assume that $\tonemom>D$ and that $B>D^6$ for some $D>0$. Then there exists a
positive constant $C$, independent of $D$, such that
\begin{equation}
\label{eq:bigf}
\q_m[\psi] \geq \bigl(\Theta_0 + C D^2 B^{-1/3}\bigr)\|\psi\|^2
\end{equation}
for all $\psi\in\dom(\q_m)$.
\end{lemma}

\begin{prf}
By the assumption on $\tonemom$ we have
\begin{equation*}
B^{-1/6}\Bigl(\tonemom+\frac{\rho^2}{2}\Bigr)> D B^{-1/6}
\end{equation*}
for all $\rho\in\mathbb{R}$. Using the de~Gennes model operator $\dG$ from
Appendix~\ref{sec:dG} we get a positive constant $C$ such that
\begin{equation*}
\begin{aligned}
\eigone{\dG}\biggl(\hmom_0+B^{-1/6}\Bigl(\tonemom+\frac{\rho^2}{2}\Bigr)\biggr)
& \geq \eigone{\dG}\bigl(\hmom_0+ D B^{-1/6}\bigr)\\ & \geq \Theta_0 + C D^2
B^{-1/3},\quad B>D^6,
\end{aligned}
\end{equation*}
from which~\eqref{eq:bigf} follows.\qed
\end{prf}

\begin{lemma}
\label{lem:rholoctwo}
Assume that $|\tonemom|\leq D_0 B^{1/6}$. Then there exist positive constants 
$C$ and $\widetilde{D}$ such that if $B>D_0^{18}$, $D>\widetilde{D}$ and 
$\tonemom<-D$ then
\begin{equation*}
\q_m[\psi] \geq \bigl(\Theta_0 + C D^{1/2} B^{-1/3}\bigr)\|\psi\|^2
\end{equation*}
for all $\psi\in\dom(\q_m)$.
\end{lemma}

\begin{prf}
We assume that $\tonemom<-D$, for some constant $D>0$. Along the proof we will 
get some constraints on $D$ that finally will determine $\widetilde{D}$. 

We first assume that $D>1$. Let $0\leq \chi_{1,B}(\rho)\leq 1$ be a smooth
cut-off function that satisfies:
\begin{itemize}
\item[(A)] $\chi_{1,B}(\rho) = 1$ if
$(1-\frac{1}{2}|\tonemom|^{-1/4})\sqrt{2|\tonemom|}\leq |\rho|\leq
(1+\frac{1}{2}|\tonemom|^{-1/4})\sqrt{2|\tonemom|}$.
\item[(B)] $\chi_{1,B}(\rho) = 0$ if $|\rho|\leq
(1-|\tonemom|^{-1/4})\sqrt{2|\tonemom|}$ or $|\rho|\geq
(1+|\tonemom|^{-1/4})\sqrt{2|\tonemom|}$.
\item[(C)] $|\chi_{1,B}'(\rho)|\leq \ccon_1|\tonemom|^{-1/4}$ for some constant 
$\ccon_1\geq0$.
\item[(D)] The function $\chi_{2,B}(\rho)=\sqrt{1-\chi_{1,B}^2(\rho)}$ satisfies
$|\chi_{2,B}'(\rho)|\leq \ccon_2|\tonemom|^{-1/4}$ for some constant 
$\ccon_2\geq 0$.
\end{itemize}
The IMS formula gives
\begin{equation*}
\q_m[\psi] = \q_m[\chi_{1,B}\psi]+\q_m[\chi_{2,B}\psi]- B^{-1/3} \sum_{j=1}^2
\int_{\mathbb{R}^2_+} \bigl|\chi_{j,B}'(\rho)\bigr|^2\cdot 
|\psi|^2 \ed\tau \ed\rho.
\end{equation*}
The error term above is bounded as
\begin{equation}
\label{eq:imstwo}
\begin{aligned}
B^{-1/3} \sum_{j=1}^2 \int_{\mathbb{R}^2_+} \bigl|\chi_{j,B}'(\rho)\bigr|^2\cdot
|\psi|^2 d \tau \ed\rho &\leq (\ccon_1^2+\ccon_2^2)|\tonemom|^{-1/2}B^{-1/3}
\|\psi\|^2\\
&\leq (\ccon_1^2+\ccon_2^2)D^{-1/2}B^{-1/3}\|\psi\|^2.
\end{aligned}
\end{equation}
From (B) we see that the support of $\chi_{1,B}$ is included in the set
\begin{equation*}
\Bigl\{\rho~\Bigm|~\tonemom+\frac{\rho^2}{2}>(-2|\tonemom|^{3/4}+|\tonemom|^{1/2})
\ \text{or}\ 
\tonemom+\frac{\rho^2}{2}< (2|\tonemom|^{3/4}+|\tonemom|^{1/2})\Bigr\},
\end{equation*}
so especially if $|\tonemom|>16$ it holds that
\begin{equation*}
\Bigl|\tonemom+\frac{\rho^2}{2}\Bigr| <  \frac{5}{2}|\tonemom|^{3/4} 
\end{equation*}
on the support of $\chi_{1,B}$. By the estimate $|\tonemom|\leq D_0B^{1/6}$ we 
have
\begin{equation*}
B^{-1/6}\Bigl|\tonemom+\frac{\rho^2}{2}\Bigr|\leq \frac{5}{2}D_0^{3/4} B^{-1/24}
\end{equation*}
on the support of $\chi_{1,B}$. If $B$ is sufficiently large ($B>D_0^{18}$) we
can Taylor expand $\eigone{\dG}$ to find a positive constant $C$ such that
\begin{equation*}
\eigone{\dG}\biggl(\hmom_0+B^{-1/6}\Bigl(\tonemom+\frac{\rho^2}{2}\Bigr)\biggr)
\geq
\Theta_0 + C \Bigl(\tonemom+\frac{\rho^2}{2}\Bigr)^2B^{-1/3}
\end{equation*}
for all $\rho$ on the support of $\chi_{1,B}$. We insert this into $\q_m$, to
get
\begin{equation*}
\begin{aligned}
\q_m[&\chi_{1,B}\psi] \\& \geq \int_{\mathbb{R}^2_+}
\Bigl[\Theta_0\bigl|\chi_{1,B}\psi\bigr|^2 +
B^{-1/3}\Bigl(|\PD{(\chi_{1,B}\psi)}{\rho}|^2+
C\Bigl(\tonemom+\frac{\rho^2}{2}\Bigr)^2|\chi_{1,B}\psi|^2 \Bigr)\Bigr]
\ed\tau \ed\rho\\
 &\geq \Bigl(\Theta_0 +
 C^{1/3}2^{-2/3}\eigone{\M}\bigl(C^{1/3}2^{1/3}\tonemom\bigr)B^{-1/3}
 \Bigr)\|\chi_{1,B}\psi\|^2,
\end{aligned}
\end{equation*}
where we have used the notation of the Montgomery model in
Appendix~\ref{sec:Mg}. By Lemma~\ref{lem:montgomery} it follows that
if $|\tonemom|$ is sufficiently large then
\begin{equation*}
\eigone{\M}\bigl(C^{1/3}2^{1/3}\tonemom\bigr) 
\geq 2 C^{1/6}2^{1/6}|\tonemom|^{1/2},
\end{equation*}
and so
\begin{equation}
\label{eq:intone}
\begin{aligned}
\q_m[\chi_{1,B}\psi] &\geq \bigl(\Theta_0 +
2^{1/2}C^{1/2}|\tonemom|^{1/2}B^{-1/3}\bigr)\|\chi_{1,B}\psi\|^2\\
&\geq \bigl(\Theta_0 + 2^{1/2}C^{1/2}D^{1/2}B^{-1/3}\bigr)\|\chi_{1,B}\psi\|^2.
\end{aligned}
\end{equation}
In the same way we show that on the support of $\chi_{2,B}$ it holds that
\begin{equation*}
\Bigl|\tonemom+\frac{\rho^2}{2}\Bigr|  > \frac{1}{2} |\tonemom|^{3/4}
\end{equation*}
if $|\tonemom|>1/16$. This implies that there exists a constant $C>0$ such that
\begin{equation}
\label{eq:inttwo}
\begin{aligned}
\q_m[\chi_{2,B}\psi] & \geq \eigone{\dG}\Bigl(\hmom_0\pm
\frac{1}{2}|\tonemom|^{3/4}B^{-1/6}\Bigr)\|\chi_{2,B}\psi\|^2\\
&\geq \bigl(\Theta_0+C |\tonemom|^{3/2}B^{-1/3}\bigr)\|\chi_{2,B}\psi\|^2\\
&\geq \bigl(\Theta_0+C D^{3/2} B^{-1/3}\bigr)\|\chi_{2,B}\psi\|^2.
\end{aligned}
\end{equation}
The proof is completed by combining the
equations~\eqref{eq:imstwo},~\eqref{eq:intone} and~\eqref{eq:inttwo} for a
sufficiently large $D$.\qed
\end{prf}

The case left to study is when there exists a constant $D$ such that
$|\tonemom|<D$. 

\begin{lemma}
\label{lem:rholocthree}
Assume that $|\tonemom|\leq D$. There exist positive constants $C_1$, $C_2$,
$\widetilde{M}$ and $B_0$ such that with $W_B$ from~\eqref{eq:WB}
\begin{equation*}
\q_m[\psi] \geq \int_{\mathbb{R}^2_+} W_B(\rho) |\psi|^2 \ed\tau \ed\rho
\end{equation*}
for all $B>B_0$ and $\psi\in\dom(\q_m)$ if $M>\widetilde{M}$.
\end{lemma}

\begin{prf}
Fix $M>0$, to be specified below. Let us introduce a smooth cut-off function
$\chi_{1,M}$ that satisfies the following properties
\begin{itemize}
\item[(i)] $0\leq \chi_{1,M}(\rho)\leq 1$
\item[(ii)] $\chi_{1,M}(\rho)=1$ if $|\rho|<M$
\item[(iii)] $\chi_{1,M}(\rho)=0$ if $|\rho|\geq 2M$
\item[(iv)] There exists a constant $\ccon_1>0$ such that $|\chi_{1,M}'(\rho)|
\leq
\ccon_1/M$ for all $\rho$.
\item[(v)] The function $\chi_{2,M}(\rho)=\sqrt{1-\chi_{1,M}^2(\rho)}$
satisfies $|\chi_{2,M}'(\rho)|\leq \ccon_2/M$ for some constant $\ccon_2>0$.
\end{itemize}
By the IMS formula, it holds that
\begin{equation*}
\q_m[\psi] = \q_m[\chi_{1,M}\psi]+\q_m[\chi_{2,M}\psi] - B^{-1/3} \sum_{j=1}^2
\int_{\mathbb{R}^2_+} \bigl|\chi_{j,M}'(\rho)\bigr|^2\cdot |\psi|^2 
\ed\tau \ed\rho.
\end{equation*}
The localization error is bounded by
\begin{equation}\label{eq:locerr}
B^{-1/3} \sum_{j=1}^2 \int_{\mathbb{R}^2_+} \bigl|\chi_{j,M}'(\rho)\bigr|^2\cdot
|\psi|^2 \ed\tau \ed\rho\leq (\ccon_1^2+\ccon_2^2)\frac{1}{M^2}B^{-1/3}
\|\psi\|^2
\end{equation}
so by choosing $M$ large, we can make this error small. We locally introduce the
notation $\psi_j=\chi_{j,M}\psi$, for $j=1,2$. On the support of $\psi_1$ we can
use the Taylor expansion of $\eigone{\dG}$ to get the existence of a constant
$C>0$ such that
\begin{equation*}
\eigone{\dG}\biggl(\hmom_0+B^{-1/6}\Bigl(\tonemom +
\frac{\rho^2}{2}\Bigr)\biggr)\geq \Theta_0 +
CB^{-1/3}\Bigl(\tonemom+\frac{\rho^2}{2}\Bigr)^2,
\end{equation*}
for $B$ large enough. This gives 
\begin{equation*}
\q_m[\psi_1] \geq \int_{\mathbb{R}^2_+}
\Theta_0|\psi_1|^2+B^{-1/3}\biggl(|\PD{\psi_1}{\rho}|^2 + C\Bigl(\tonemom
+\frac{\rho^2}{2}\Bigr)^2|\psi_1|^2 \biggr)\ed\tau \ed\rho.
\end{equation*}
Next, we use the modified Montgomery operator in
Appendix~\ref{sec:Mg}, to get
\begin{equation*}
\q_m[\psi_1] \geq \int_{\mathbb{R}^2_+} \Theta_0|\psi_1|^2 + 
B^{-1/3}C^{1/3}2^{-2/3}
\eigone{\M}\bigl(C^{1/3}2^{1/3}\tonemom\bigr)
|\psi_1|^2 \ed\tau \ed\rho.
\end{equation*}
Since $\eigone{\M}$ has minimum $\hat{\nu}_0$ we get a constant
$C_1=\frac{1}{2}C^{1/3}2^{-2/3}\hat{\nu}_0$ such that
\begin{equation*}
\q_m[\psi_1] \geq \bigl(\Theta_0 + 2C_1 B^{-1/3}\bigr)\|\psi_1\|^2
\end{equation*}
On the support of $\psi_2$ we have
\begin{equation*}
\Bigl|\tonemom+\frac{\rho^2}{2}\Bigr| \geq
\frac{M^2}{8}-D \geq \frac{M^2}{10}
\end{equation*}
where $\widetilde{M}$ is chosen large enough, so that the last inequality holds
for $M>\widetilde{M}$. There exists a constant $C_2>0$, independent of $M$, such
that
\begin{equation*}
\begin{aligned}
\eigone{\dG}\biggl(\hmom_0 +
B^{-1/6}\Bigl(\tonemom+\frac{\rho^2}{2}\Bigr)\biggr) & \geq
\eigone{\dG}\bigl(\hmom_0 \pm B^{-1/6}M^2/10\bigr)\\
&\geq \Theta_0 + 2C_2M^4 B^{-1/3}
\end{aligned}
\end{equation*}
and so we get
\begin{equation*}
\q_m[\psi_2] \geq \bigl(\Theta_0 + 2C_2 M^4 B^{-1/3}\bigr)\|\psi_2\|^2.
\end{equation*}
This finishes the proof if we choose $\widetilde{M}$ so large that the 
localization error~\eqref{eq:locerr} is dominated by 
$C_1B^{-1/3}\|\psi_1\|^2 + C_2 M^4 B^{-1/3}\|\psi_2\|^2$.\qed
\end{prf}

%
%
%
%
%
%            LOCALIZATIONS
%
%
%
%
%

\section{An improved localization formula}\label{sec:newloc}

\begin{prop}
\label{prop:rhoestright}
Let $\omega > 0$ and $a>0$. Then there exist positive constants $B_0$ and $C_0$
such that if $B>B_0$ and $\psi$ satisfy $\Q_m(B)\psi = \lambda\psi$ with 
$\lambda \leq \Theta_0+\omega B^{-1/3}$ then
\begin{equation}
\label{eq:agmon}
\int_{\mathbb{R}^2_+}
e^{2a|\rho|}\bigl(|\psi|^2+|\PD{\psi}{\tau}|^2+B^{-1/3}|\PD{\psi}{\rho}|^2\bigr)
\ed\tau \ed\rho \leq C_0 \|\psi\|^2.
\end{equation}
\end{prop}

\begin{prf}
Let $\chi_{1,M}$ and $\chi_{2,M}$ be the cut-off functions from the proof of
Lemma~\ref{lem:rholocthree}, where $M$ is going to be specified below. Also, for
$\epsilon>0$, let
\begin{equation}\label{eq:veps}
v_\epsilon(\rho) = \frac{|\rho|}{1+\epsilon|\rho|}.
\end{equation}
This function $v_\epsilon$ is bounded and continuous on $\mathbb{R}$ and
differentiable everywhere except at $0$. Moreover $|v'_\epsilon(\rho)|\leq 1$
for all $\rho\neq0$. In particular the function $\chi_{2,M}\psi
e^{a v_\epsilon(\rho)}$ belongs to the domain of $\q_m$. We use integration by
parts (the IMS formula) to get
\begin{equation}
\label{eq:agmonloc}
\lambda \|\chi_{2,M}\psi e^{a v_\epsilon(\rho)}\|^2 = \q_m\bigl[\chi_{2,M}\psi
e^{a v_\epsilon(\rho)}\bigr] - \int_{\mathbb{R}^2_+}
\bigl|\PD{}{\rho}\bigl(\chi_{2,M}e^{a v_\epsilon(\rho)}\bigr)\psi\bigr|^2 
\ed\tau \ed\rho.
\end{equation}
Next, we choose $M$ such that both $\bigl(C_2M^4-\omega-2a^2\bigr)\geq 1$ and
$M>\widetilde{M}$ hold, where $\widetilde{M}$ is the constant in
Proposition~\ref{prop:rholoc}, and we choose $B$ greater than the constant $B_0$ 
in Proposition~\ref{prop:rholoc}. Using the assumption on $\lambda$ and the 
lower bound on $\q_m$ from Lemma~\ref{lem:rholocthree} we get
\begin{multline}\label{eq:agmonone}
\int_{\mathbb{R}^2_+}\bigl(C_2M^4-\omega-2a^2
v_\epsilon'(\rho)^2\bigr)\bigl|\chi_{2,M}\psi e^{a v_\epsilon(\rho)}\bigr|^2
\ed\tau \ed\rho \\
\leq 2 \int_{\mathbb{R}^2_+} |\chi_{2,M}'(\rho)|^2 
e^{2a v_\epsilon(\rho)}|\psi|^2 \ed\tau \ed\rho.
\end{multline}
The function $\chi_{2,M}'(\rho)$ is supported in the set $\{\rho\in\mathbb{R}
\mid M<\rho<2M\}$, where also the inequality 
$e^{a v_\epsilon(\rho)}\leq e^{4aM}$ holds. Inserting this and the choice of 
$M$ in~\eqref{eq:agmonone} we get
\begin{equation}
\label{eq:agmonouteps}
\int_{\mathbb{R}^2_+}\bigl|\chi_{2,M}\psi e^{a v_\epsilon(\rho)}\bigr|^2 
\ed\tau \ed\rho \leq 2e^{4aM}\int_{\mathbb{R}^2_+}|\psi|^2 \ed\tau \ed\rho.
\end{equation}
Since the right-hand side is independent of $\epsilon$ we can let $\epsilon$
tend to zero and use monotone convergence to get
\begin{equation}
\label{eq:agmonout}
\int_{\mathbb{R}^2_+} \chi_{2,M}^2 e^{2a |\rho|}|\psi|^2 \ed\tau \ed\rho \leq
2e^{4aM}\int_{\mathbb{R}^2_+}|\psi|^2 \ed\tau \ed\rho.
\end{equation}
Since $\chi_{1,M}$ is supported in 
$\bigl\{\rho\in\mathbb{R}~\bigm|~|\rho|<2M\bigr\}$ we have
\begin{equation}
\label{eq:agmonin}
\int_{\mathbb{R}^2_+} \chi_{1,M}^2 e^{2a |\rho|}|\psi|^2 \ed\tau \ed\rho \leq
e^{4aM}\int_{\mathbb{R}^2_+}|\psi|^2 \ed\tau \ed\rho.
\end{equation}
Combining~\eqref{eq:agmonout} and~\eqref{eq:agmonin} gives the $L^2$-bound
in~\eqref{eq:agmon}. Next we turn to the terms involving derivatives. 
Using the triangle inequality, we have for the $\rho$-derivative
\begin{multline*}
\int_{\mathbb{R}^2_+}e^{2av_\epsilon(\rho)}\chi_{2,M}^2 |\PD{}{\rho}\psi|^2
\ed\tau \ed\rho \\
 \leq 2 \int_{\mathbb{R}^2_+}
\bigl|\PD{}{\rho}\bigl(e^{av_\epsilon(\rho)}\chi_{2,M} \psi\bigr)\bigr|^2 
\ed\tau\ed\rho + 
2\int_{\mathbb{R}^2_+}
\bigl|\PD{}{\rho}\bigl(e^{av_\epsilon(\rho)}\chi_{2,M}\bigr)\psi\bigr|^2 
\ed\tau \ed\rho\\
\leq 2 B^{1/3}\q_m\bigl[e^{av_\epsilon(\rho)}\chi_{2,M} \psi\bigr]+
2\int_{\mathbb{R}^2_+}
\bigl|\PD{}{\rho}\bigl(e^{av_\epsilon(\rho)}\chi_{2,M}\bigr)\psi\bigr|^2
\ed\tau \ed\rho.
\end{multline*}
The corresponding inequality for the $\tau$-derivative is, since $v_\epsilon$ 
and $\chi_{2,M}$ are independent of $\tau$,
\begin{equation*}
\int_{\mathbb{R}^2_+}e^{2av_\epsilon(\rho)}\chi_{2,M}^2 |\PD{}{\tau}\psi|^2
\ed\tau \ed\rho \leq \q_m\bigl[e^{av_\epsilon(\rho)}\chi_{2,M} \psi\bigr].
\end{equation*}
Combining these two inequalities with~\eqref{eq:agmonloc} gives
\begin{multline*}
\int_{\mathbb{R}^2_+}e^{2av_\epsilon(\rho)}\chi_{2,M}^2
\bigl(|\PD{}{\tau}\psi|^2+ B^{-1/3}|\PD{}{\rho}\psi|^2\bigr) \ed\tau \ed\rho \\
\leq 2B^{-1/3}\int_{\mathbb{R}^2_+}\bigl|\PD{}{\rho}
\bigl(e^{av_\epsilon(\rho)}\chi_{2,M}\bigr)\psi\bigr|^2 \ed\tau \ed\rho
+ 3\q_m\bigl[e^{av_\epsilon(\rho)}\chi_{2,M} \psi\bigr]\\
\leq (2B^{-1/3}+3)\int_{\mathbb{R}^2_+}
\bigl|\PD{}{\rho}\bigl(e^{av_\epsilon(\rho)}\chi_{2,M}\bigr) \psi\bigr|^2 
\ed\tau \ed\rho+3\lambda \|\chi_{2,M}\psi e^{a v_\epsilon(\rho)}\|^2.
\end{multline*}
Moreover,
\begin{equation*}
\bigl|\PD{}{\rho}\bigl(e^{av_\epsilon(\rho)}\chi_{2,M}\bigr)\bigr| =
|av_\epsilon'(\rho)+\chi_{2,M}'(\rho)|e^{av_\epsilon(\rho)}\leq
(a+\ccon_2/M)e^{av_\epsilon(\rho)},
\end{equation*}
so for $B>1$ we can use~\eqref{eq:agmonouteps} to get
\begin{multline*}
\int_{\mathbb{R}^2_+}e^{2av_\epsilon(\rho)}\chi_{2,M}^2
\bigl(|\PD{}{\tau}\psi|^2+ B^{-1/3}|\PD{}{\rho}\psi|^2\bigr) \ed\tau \ed\rho \\
 \leq \bigl(5(a+\ccon_2/M)+3(\Theta_0+\omega)\bigr)\|\chi_{2,M}\psi e^{a
v_\epsilon(\rho)}\|^2 \\
 \leq 2\bigl[5(a+\ccon_2/M)+3(\Theta_0+\omega)\bigr]e^{4aM}
\int_{\mathbb{R}^2_+}|\psi|^2 \ed\tau \ed\rho.
\end{multline*}
By monotone convergence we have
\begin{multline}\label{eq:newagmonout}
\int_{\mathbb{R}^2_+}e^{2a|\rho|}\chi_{2,M}^2 \bigl(|\PD{}{\tau}\psi|^2+
B^{-1/3}|\PD{}{\rho}\psi|^2\bigr) \ed\tau \ed\rho \\
\leq
2\bigl[5(a+\ccon_2/M)+3(\Theta_0+\omega)\bigr]e^{4aM} \int_{\mathbb{R}^2_+}
|\psi|^2
\ed\tau \ed\rho.
\end{multline}
The same estimate with $\chi_{1,M}$ in place of $\chi_{2,M}$ is easier since we
do not have to use $v_\epsilon$ and the functions involved have compact support.
The result is
\begin{multline}\label{eq:newagmonin}
\int_{\mathbb{R}^2_+}e^{2a|\rho|}\chi_{1,M}^2 \bigl(|\PD{}{\tau}\psi|^2+
B^{-1/3}|\PD{}{\rho}\psi|^2\bigr) \ed\tau \ed\rho \\
\leq
2\bigl[5(a+\ccon_1/M)+3(\Theta_0+\omega)\bigr]e^{4aM} \int_{\mathbb{R}^2_+}
|\psi|^2 \ed\tau \ed\rho.
\end{multline}
Finally, a combination of the equations
\eqref{eq:agmonout},~\eqref{eq:agmonin},~\eqref{eq:newagmonout}
and~\eqref{eq:newagmonin} implies~\eqref{eq:agmon}. 
\qed
\end{prf}

\begin{cor}
\label{cor:rhoestright}
For all $\omega>0$ and $n\in\mathbb{N}$ there exist positive constants 
$B_n$ and $C_n$ such that if
$B>B_n$ and $\psi$ satisfies $\Q_m\psi = \lambda\psi$ with 
$\lambda \leq \Theta_0+\omega B^{-1/3}$ then
\begin{equation*}
\int_{\mathbb{R}^2_+}
|\rho|^{n}\bigl(|\psi|^2+|\PD{}{\tau}\psi|^2+B^{-1/3}|\PD{}{\rho}\psi|^2\bigr)
\ed\tau \ed\rho \leq C_n \|\psi\|^2.
\end{equation*}
\end{cor}

\begin{prf}
This follows from Proposition~\ref{prop:rhoestright}, by noting that all the
terms in the Taylor expansion of $e^{2a|\rho|}$ are positive, and thus for all
non-negative integers $n$ it holds that $e^{2a|\rho|}\geq (2a|\rho|)^n/n!$.\qed
\end{prf}

%
%
%
%
%
%            SPECTRAL GAP
%
%
%
%
%

\section{Improved lower bounds and a spectral gap}\label{sec:gap}

\begin{prop}\label{prop:gap}
Let $\tonemom$ be as in~\eqref{eq:tonemom}, i.e., 
$\tonemom = \bigl(\frac{1}{\sqrt{B}}(m-B/2)-\hmom_0\bigr)B^{1/6}$ and 
$\eigone{\widetilde{M}}$ be the lowest eigenvalue of the Montgomery operator,
see Appendix~\ref{sec:Mg}.
\begin{itemize}
\item[(A)] For all $C_1>0$ there exist constants $B_0$ and $C_0$ 
(independent of $m$) such that if $|\tonemom|<C_1$ and $B>B_0$ then
\begin{equation}\label{eq:proplow}
\q_m[\psi]\geq 
\bigl(\Theta_0 + \eigone{\widetilde{M}}(\tonemom)B^{-1/3}-C_0 B^{-3/8}\bigr)
\|\psi\|^2
\end{equation}
for all $\psi\in\dom(\Q_m(B))$.
\item[(B)] For all $C_1>0$ there exist positive constants $\gamma$, $B_0$ and
$C_0$ (independent of $m$) such that if $|\tonemom|<C_1$ it holds that
\begin{equation}\label{eq:propgap}
\eig{2}{\Ham_m(B)} \geq \Theta_0 B+
(\widehat{\gamma}_0+\gamma) B^{2/3}-C_0B^{7/12}
\end{equation}
if $B>B_0$. In particular, if $\tonemom$ is bounded there exists a positive 
constant $B_1$ (independent of $m$), for $B>B_1$ then the set
\begin{equation*}
\spec\bigl(\Ham_m(B)\bigr)\cap 
(-\infty,\Theta_0 B+ (\widehat{\gamma}_0+\gamma/2)B^{2/3})
\end{equation*}
is either empty or consists of the lowest eigenvalue of $\Ham_m(B)$.
\end{itemize}
\end{prop}

\begin{prf}
 We recall that
\begin{equation*}
\q_m[\psi] = \int_{\mathbb{R}^2_+} \bigl|\PD{\psi}{\tau}\bigr|^2 +
\biggl(\tau+\hmom_0+B^{-1/6}\Bigl(\tonemom+\frac{\rho^2}{2}\Bigr)\biggr)^2
|\psi|^2 + B^{-1/3}\bigl|\PD{\psi}{\rho}\bigr|^2 \ed\tau \ed\rho.
\end{equation*}
We start with the proof of~(A).
Fix $0<\cpar<1/12$. Let us introduce a smooth cut-off 
function $0\leq \chi_{1,B}(\rho)\leq 1$ that satisfies the following properties
\begin{itemize}
\item[(i)] $\chi_{1,B}(\rho)=1$ if $|\rho|<B^\cpar$
\item[(ii)] $\chi_{1,B}(\rho)=0$ if $|\rho|\geq 2B^\cpar$
\item[(iii)] There exists a constant $\ccon_1>0$ such that 
$|\chi_{1,B}'(\rho)|\leq \ccon_1 B^{-\cpar}$ for all $\rho$.
\item[(iv)] The function $\chi_{2,B}(\rho)=\sqrt{1-\chi_{1,B}^2(\rho)}$
satisfies $|\chi_{2,B}'(\rho)|\leq \ccon_2 B^{-\cpar}$ for some constant 
$\ccon_2>0$.
\end{itemize}
We denote by $\psi_j=\chi_{j,B}\psi$. Then clearly both $\psi_1$ and $\psi_2$ 
belong to the domain of $\q_m$ and by the IMS formula
\begin{equation*}
\q_m[\psi] = \q_m[\psi_1]+\q_m[\psi_2] - B^{-1/3} \sum_{j=1}^2
\int_{\mathbb{R}^2_+} \bigl|\chi_{j,B}'(\rho)\bigr|^2\cdot 
|\psi|^2 d\tau d\rho.
\end{equation*}
The IMS error is easily seen to be bounded from below by some negative constant 
times $B^{-1/3-2\cpar}\|\psi\|^2$. By Proposition~\ref{prop:rholoc}
\begin{equation*}
\q_m[\psi_2]\geq (\Theta_0 + C_0B^{-1/3})\|\psi_2\|^2, 
\end{equation*}
where we can make the constant $C_0$ as large as we want, by choosing $B$ large
(using the properties of the support of $\chi_{2,B}$). 

We turn to $\q_m[\psi_1]$, and note that, with $\eigone{\dG}$ from
Appendix~\ref{sec:dG},
\begin{equation*}
\q_m[\psi_1]\geq \int_{\mathbb{R}^2_+} 
\eigone{\dG}\biggl(\hmom_0+B^{-1/6}
\Bigl(\tonemom+\frac{\rho^2}{2}\Bigr)\biggr)
|\psi_1|^2 + B^{-1/3}\bigl|\PD{\psi_1}{\rho}\bigr|^2 \ed\tau \ed\rho.
\end{equation*}
Since $\cpar<1/12$ we can Taylor expand the first eigenvalue $\eigone{\dG}$ to 
get a constant $C>0$ such that, on the support of $\psi_1$,
\begin{multline*}
\eigone{\dG}
\biggl(\hmom_0+B^{-1/6}\Bigl(\tonemom+\frac{\rho^2}{2}\Bigr)\biggr)\\
\geq \eigone{\dG}(\hmom_0) 
+\frac{1}{2}\eigone{\dG}''(\hmom_0)\Bigl(\tonemom+\frac{\rho^2}{2}\Bigr)^2 
B^{-1/3}
-C \Bigl(\tonemom+\frac{\rho^2}{2}\Bigr)^3 B^{-1/2}\\
\geq \Theta_0 + \dGdiff\Bigl(\tonemom+\frac{\rho^2}{2}\Bigr)^2 B^{-1/3} 
- CB^{-1/2+6\cpar}
\end{multline*}
We insert this into $\q_m$ to get
\begin{equation*}
\q_m[\psi_1]\geq \Theta_0\|\psi_1\|^2 + 
B^{-1/3}\int_{\mathbb{R}^2_+} \bigl|\PD{\psi_1}{\rho}\bigr|^2 
+ \dGdiff \Bigl(\tonemom+\frac{\rho^2}{2}\Bigr)^2|\psi_1|^2
 \ed\tau \ed\rho - C B^{-1/2+6\cpar}\|\psi_1\|^2.
\end{equation*}
Next, we use the modified Montgomery model from Appendix~\ref{sec:Mg} to 
estimate the integral above,
\begin{equation*}
\int_{\mathbb{R}^2_+} \bigl|\PD{\psi_1}{\rho}\bigr|^2 + 
\dGdiff \Bigl(\tonemom+\frac{\rho^2}{2}\Bigr)^2|\psi_1|^2
 \ed\tau \ed\rho \geq \eigone{\widetilde{\M}}(\tonemom)\|\psi_1\|^2.
\end{equation*}
We choose $\cpar=1/48$ and put the pieces together to obtain~\eqref{eq:proplow}.

We continue with the proof of~(B).
It is enough to prove~\eqref{eq:propgap} for $\Q_m(B)$, i.e.,
\begin{equation*}
\eig{2}{\Q_m(B)} \geq \Theta_0 B+
(\widehat{\gamma}_0+\gamma) B^{2/3}+\bigo(B^{7/12}),\quad \text{as }B\to\infty,
\end{equation*}
The inequality for 
$\Ham_m(B)$ is then a direct consequence of Lemmas~\ref{lem:psicutoff} 
and~\ref{lem:bhalf}.

Let $\psi^{(1)}$ and $\psi^{(2)}$ denote the first two normalized eigenfunctions
of $\Q_m(B)$ 
and assume that $m$ is such that
$\eig{2}{\Q_m(B)}\leq \Theta_0 + \omega B^{-1/3}$ for some $\omega>0$, otherwise
there is nothing to prove.

We use the same cut-off function $0\leq \chi_{1,B}(\rho)\leq 1$ as in the 
proof of~(A), but with $\cpar=1/72$.
We also introduce the quadratic form $\q_m^D$ with the same action as $\q_m$, 
but with an additional Dirichlet 
condition at $|\rho|=2B^\cpar$. For simplicity, we extend functions in the domain
of $\q_m^D$ by zero for $|\rho|>2B^\cpar$. We also denote by $\Q_m^D(B)$ the 
corresponding self-adjoint operator.

We start by showing that
\begin{equation}
\label{eq:gapone}
\eig{2}{\Q_m(B)} \geq \eig{2}{\Q_m^D(B)}+\bigo(B^{-\infty})
,\quad \text{as }B\to\infty.
\end{equation}
Let us write $\psi^{(k)}_j=\chi_{j,B}\psi^{(k)}$, $j,k=1,2$. 
By the IMS formula, it holds that
\begin{equation*}
\q_m[\psi^{(k)}] = \q_m[\psi^{(k)}_1]+\q_m[\psi^{(k)}_2] - B^{-1/3} \sum_{j=1}^2
\int_{\mathbb{R}^2_+} \bigl|\chi_{j,B}'(\rho)\bigr|^2\cdot 
|\psi^{(k)}|^2 d\tau d\rho,\quad k=1,2.
\end{equation*}
By Proposition~\ref{prop:rhoestright}, we have for $k=1,2$, as $B\to\infty$,
\begin{gather}
\label{eq:smallterms}
\|\psi^{(k)}_2\|=\bigo(B^{-\infty})\|\psi^{(k)}\|,\quad
\q_m[\psi^{(k)}_2]=\bigo(B^{-\infty})\|\psi^{(k)}\|^2,\quad \text{and}\\
B^{-1/3} \sum_{j=1}^2 \int_{\mathbb{R}^2_+} \bigl|\chi_{j,B}'(\rho)\bigr|^2\cdot
|\psi^{(k)}|^2 \ed\tau \ed\rho  = \bigo(B^{-\infty})\|\psi^{(k)}\|^2.\notag
\end{gather}
By the min-max principle we have
\begin{equation}
\label{eq:minmaxone}
\eig{2}{\Q_m(B)} = \max_{\psi\in\Span\{\psi^{(1)},\psi^{(2)}\}} 
\frac{\q_m[\psi]}{\|\psi\|^2} 
= \max_{\substack{\alpha,\beta\\ |\alpha|^2+|\beta|^2=1}} 
\q_m[\alpha\psi^{(1)}+\beta\psi^{(2)}].
\end{equation}
Using Proposition~\ref{prop:rhoestright} and~\eqref{eq:smallterms} we see that
\begin{equation*}
\q_m[\alpha\psi^{(1)}+\beta\psi^{(2)}] = 
\q_m[\alpha\psi^{(1)}_1+\beta\psi^{(2)}_1] + \bigo(B^{-\infty}),
\quad \text{as }B\to\infty.
\end{equation*}
In the right-hand side we can write $\q_m^D$ instead of $\q_m$. It follows
from~\eqref{eq:smallterms} that 
\begin{equation*}
\|\alpha \psi^{(1)}_1 + \beta \psi^{(2)}_1\|^2=1+\bigo(B^{-\infty}),
\quad \text{as }B\to\infty. 
\end{equation*}
Using the
min-max principle for $\q_m^D$ we have
\begin{align*}
\eig{2}{\Q_m^D(B)} &= \min_{\dim V = 2} \max_{\psi\in V} 
\frac{\q_m^D[\psi]}{\|\psi\|^2}
\leq \max_{\substack{\alpha,\beta\\ |\alpha|^2+|\beta|^2=1}} 
\frac{\q_m[\alpha\psi_1^{(1)}+\beta\psi_1^{(2)}]}
{\|\alpha\psi_1^{(1)}+\beta\psi_1^{(2)}\|^2}\\
&= \max_{\substack{\alpha,\beta\\ |\alpha|^2+|\beta|^2=1}} 
\q_m[\alpha\psi_1^{(1)}+\beta\psi_1^{(2)}] + \bigo(B^{-\infty})
,\quad \text{as }B\to\infty.
\end{align*}
Combining this with~\eqref{eq:minmaxone} we get~\eqref{eq:gapone}.

Next, we show the existence of a positive constant $\gamma$ such that the
inequality
\begin{equation}
\label{eq:gaptwo}
\eig{2}{\Q_m^D(B)} \geq \Theta_0 
+\bigl(\widehat{\gamma}_0 + \gamma\big)B^{-1/3}+\bigo(B^{-5/12})
,\quad \text{as }B\to\infty,
\end{equation}
holds for all $m\in\mathbb{Z}$ for which $|\tmom_1|$ is bounded. Let 
$\epsilon=B^{-1/3}$ and $\psi\in\dom(\q_m^D)$. We write $\q_m^D[\psi]$ as
\begin{multline*}
\q_m^D[\psi] = \epsilon\int_{\mathbb{R}^2_+} \left|\PD{\psi}{\tau}\right|^2
+\bigl(\tau+\hmom_0)^2|\psi|^2 \ed\tau \ed\rho\\
           + (1-\epsilon)\int_{\mathbb{R}^2_+}
\left|\PD{\psi}{\tau}\right|^2+\left(\tau+\hmom_0 +
\frac{B^{-1/6}}{1-\epsilon}\Bigl(\tonemom+\frac{\rho^2}{2}\Bigr)
\right)^2|\psi|^2
           + \frac{B^{-1/3}}{1-\epsilon}\left|\PD{\psi}{\rho}\right|^2 
d \tau \ed\rho\\
           -\frac{\epsilon}{1-\epsilon} \int_{\mathbb{R}^2_+}
\biggl(B^{-1/6}\Bigl(\tonemom+\frac{\rho^2}{2}\Bigr)
\biggr)^2|\psi|^2 \ed\tau \ed\rho.
\end{multline*} 
For sufficiently large $B$ we use the support of $\psi$ and the assumption 
$|\tonemom|\leq C_1$ to bound the last integral, uniformly in $m$,
\begin{equation*}
\left|\frac{\epsilon}{1-\epsilon}\int_{\mathbb{R}^2_+}\left(B^{-1/6}
\Bigl(\tonemom+\frac{\rho^2}{2}\Bigr)\right)^2|\psi|^2
\ed\tau \ed\rho\right|= \bigo\bigl(B^{-2/3+4\cpar}\bigr)\|\psi\|^2
,\quad \text{as }B\to\infty.
\end{equation*}
We get that $\q_m^D[\psi]$ satisfies
\begin{equation*}
\begin{aligned}
\q_m^D[\psi] &\geq \epsilon\int_{\mathbb{R}^2_+}
\left|\PD{\psi}{\tau}\right|^2 +\bigl(\tau+\hmom_0)^2|\psi|^2 \ed\tau \ed\rho\\
           &\quad + (1-\epsilon)\int_{\mathbb{R}^2_+}
\eigone{\dG}\left(\hmom_0+\frac{B^{-1/6}}{1-\epsilon}
\Bigl(\tonemom+\frac{\rho^2}{2}\Bigr)\right)|\psi|^2
+ \frac{B^{-1/3}}{1-\epsilon}\left|\PD{\psi}{\rho}\right|^2 \ed\tau
\ed\rho \\
&\quad+\bigo\bigl(B^{-2/3+4\cpar}\bigr)\|\psi\|^2
,\quad \text{as }B\to\infty,
\end{aligned}
\end{equation*}
where $\eigone{\dG}$ is the lowest eigenvalue of the de~Gennes model, see
Appendix~\ref{sec:dG}.
We use that $\psi$ has bounded support and estimate, 
using the Taylor expansion of $\eigone{\dG}$, as $B\to\infty$,
\begin{multline*}
\int_{\mathbb{R}^2_+} \eigone{\dG}\left(\hmom_0+\frac{B^{-1/6}}{1-\epsilon}
\Bigl(\tonemom+\frac{\rho^2}{2}\Bigr)\right) |\psi|^2\ed\tau\ed\rho \\
 \geq \int_{\mathbb{R}^2_+} \biggl(\Theta_0 +
\frac{\dGdiff}{(1-\epsilon)^2}B^{-1/3}
\Bigl(\tonemom+\frac{\rho^2}{2}\Bigr)^2 \biggr)|\psi|^2 \ed\tau \ed\rho
+\bigo(B^{-1/2+6\cpar})\|\psi\|^2\\
 \geq \int_{\mathbb{R}^2_+} \biggl(\Theta_0 +
\frac{\dGdiff}{(1-\epsilon)}B^{-1/3}
\Bigl(\tonemom+\frac{\rho^2}{2}\Bigr)^2 \biggr)|\psi|^2 \ed\tau \ed\rho
+\bigo(B^{-1/2+6\cpar})\|\psi\|^2.
\end{multline*}
In the last inequality we also used that 
$(1-\epsilon)^{-2}-(1-\epsilon)^{-1}=\bigo(B^{-1/3})$ together with 
Corollary~\ref{cor:rhoestright}. Inserting in $\q_m^D$ 
we have, with the choice $\cpar=1/72$, as $B\to\infty$,
\begin{align}\label{eq:minmaxexpr}
\q_m^D[\psi] &\geq \epsilon\int_{\mathbb{R}^2_+}
\left|\PD{\psi}{\tau}\right|^2 +\bigl(\tau+\hmom_0)^2|\psi|^2 \ed\tau \ed\rho +
(1-\epsilon) \Theta_0\int_{\mathbb{R}^2_+} |\psi|^2 \ed\tau \ed\rho\\
         &\quad+ \frac{B^{-1/3}}{1-\epsilon}\int_{\mathbb{R}^2_+}
         \left|\PD{\psi}{\rho}\right|^2
         +\dGdiff\Bigl(\tonemom+\frac{\rho^2}{2}\Bigr)^2|\psi|^2
         \ed\tau \ed\rho + \bigo(B^{-5/12})\|\psi\|^2\notag\\
& = \epsilon\bigl(\dG(\hmom_0)\otimes 1\bigr)[\psi] 
+ (1-\epsilon)\Theta_0 \|\psi\|^2
+\frac{\epsilon}{1-\epsilon} 
\bigl(1\otimes \widetilde{\M}(\tonemom)\bigr)[\psi]+\bigo(B^{-5/12}).\notag
\end{align}
Here the operators $\dG$ and $\widetilde{\M}$ were introduced in 
Appendix~\ref{sec:modeloperators}. We note that the variables $\tau$ and 
$\rho$ are separated in the last expression, so if we denote by $T$ the 
operator corresponding to the form on the right-hand side above then we have
\begin{equation*}
\begin{aligned}
\eigone{T} &= \epsilon\eigone{\dG}(\hmom_0)+(1-\epsilon)\Theta_0 
+ \eigone{\widetilde{M}}(\tonemom)B^{-1/3} + \bigo(B^{-5/12})\\
&= \Theta_0 + \eigone{\widetilde{M}}(\tonemom)B^{-1/3} 
+ \bigo(B^{-5/12})
,\quad \text{as }B\to\infty.
\end{aligned}
\end{equation*}
Denote by 
\begin{equation*}
\gamma_{\dG}= \eig{2}{\dG}(\hmom_0)-\eigone{\dG}(\hmom_0) \quad 
\text{and}\quad 
\gamma_{\widetilde{\M}}= 
\inf_{|\tonemom|\leq C_1} \Bigl(\eig{2}{\widetilde{\M}}(\tonemom)
-\eigone{\widetilde{\M}}(\tonemom)\Bigr)
\end{equation*}
the spectral gaps for the de Gennes model and Montgomery model
respectively. Since both $\eigone{\dG}(\hmom_0)$ and 
$\eigone{\widetilde{\M}}(\tonemom)$ are simple eigenvalues, and $\tonemom$ is 
varying in a compact set, it follows that 
both $\gamma_{\dG}$ and $\gamma_{\widetilde{\M}}$ are strictly positive. 

For the second eigenvalue of $T$ we get
\begin{equation*}
\eig{2}{T} = \Theta_0 + \bigl(\eigone{\widetilde{M}}(\tonemom) +
\min(\gamma_{\dG},\gamma_{\widetilde{M}})\bigr)B^{-1/3}+\bigo(B^{-5/12})
,\quad \text{as }B\to\infty.
\end{equation*}
If we choose
$\gamma=\frac{1}{2}\min(\gamma_{\dG},\gamma_{\widetilde{\M}})$, 
we see that 
\begin{multline}
\label{eq:Ttwo}
\eig{2}{T}  
\geq \Theta_0 +
\bigl(\eigone{\widetilde{M}}(\tonemom) +\gamma\bigr)
B^{-1/3} +\bigo(B^{-5/12})\\
\geq \Theta_0 +
\bigl(\widehat{\gamma}_0 +\gamma\bigr)
B^{-1/3}+\bigo(B^{-5/12})
,\quad \text{as }B\to\infty,
\end{multline}
where in the last inequality we use that $\eigone{\widetilde{M}}(\tonemom) 
\geq \eigone{\widetilde{M}}(\tmom)
=\widehat{\gamma}_0$. By~\eqref{eq:minmaxexpr} it follows that
$\eig{j}{\Q_m^D(B)}\geq \eig{j}{T}$ for all $j$, so by~\eqref{eq:Ttwo} we 
get~\eqref{eq:gaptwo}.

The proof of~\eqref{eq:propgap} for $\Q_m(B)$ now follows by 
combining~\eqref{eq:gapone} and~\eqref{eq:gaptwo}.
\qed
\end{prf}

%
%
%
%
%
%            UPPER        BOUNDS
%
%
%
%
%

\section{Calculating good trial states}
\label{sec:upper}

\subsection{Statement} We provide three different estimates of 
(the lowest point in) the spectrum of $\Ham_m(B)$. Each is superior to the
others in a specific parameter regime.

When combined with Proposition~\ref{prop:gap}, Theorem~\ref{thm:goodtrial} will
give two-sided bounds on the ground state energy $\eigone{\Ham_m(B)}$.
\begin{thm}\label{thm:goodtrial}
Let $\dGdiff>0$ be the constant from Lemma~\ref{lem:thetazero}.
There exist constants $\hmom_0$, $\hmom_1$, $\hmom_2$, 
$\lambda_j$, $j=0,\ldots,5$, and polynomials 
$\lambda_4(\para)=\dGdiff\para^2+\lambda_4$, 
$\lambda_5(\para)$ with constant term $\lambda_5$, and
$\lambda_6(\mom_3)$ quadratic with quadratic
term $\dGdiff\mom_3^2$, such that with
\begin{align*}
\para &= \Bigl(m-\frac{B}{2}-\hmom_0\sqrt{B}-\hmom_1 B^{1/3}\Bigr)B^{-1/6}
-\hmom_2,\quad \text{and}\\
\mom_3&= m-\frac{B}{2}-\hmom_0\sqrt{B}-\hmom_1 B^{1/3}-\hmom_2 B^{1/6},
\end{align*}
the following holds:
\begin{itemize}
\item[(i)] For any $\widetilde{\k}_{2/6}>0$ there exist 
$\widehat{\k}_{2/6}$ and $\widehat{B}_{2/6}$ such that if 
$|\para|<\widetilde{\k}_{2/6} B^{1/6}$ 
and $B>\widehat{B}_{2/6}$ then
\begin{equation}\label{eq:goodfour}
\dist\biggl[\sum_{j=0}^{3}\lambda_j B^{-j/6} 
+\lambda_4(\para)B^{-4/6},\spec\Bigl(\frac{1}{B}\Ham_m(B)\Bigr)\biggr]
\leq\widehat{\k}_{2/6}(1+|\para|^2)B^{-5/6}.
\end{equation}
\item[(ii)] For any $\widetilde{\k}_{1/6}>0$ there exist 
$\widehat{\k}_{1/6}$ and $\widehat{B}_{1/6}$ such that if 
$|\para|<\widetilde{\k}_{1/6}$ 
and $B>\widehat{B}_{1/6}$ then
\begin{equation}\label{eq:goodfive}
\dist\biggl[\sum_{j=0}^{3}\lambda_j B^{-j/6} 
+\lambda_4(\para)B^{-4/6}+\lambda_5(\para) B^{-5/6}
,\spec\Bigl(\frac{1}{B}\Ham_m(B)\Bigr)\biggr]
\leq\widehat{\k}_{1/6}B^{-6/6}.
\end{equation}
\item[(iii)] For any $\widetilde{\k}_0>0$ there exist 
$\widehat{\k}_{0}$ and $\widehat{B}_{0}$ such that
if $|\mom_3|\leq \widetilde{\k}_{0}$ and $B>\widehat{B}_0$ then
\begin{equation}\label{eq:goodtrial}
\dist\biggl[\sum_{j=0}^{5}\lambda_j B^{-j/6}+\lambda_6(\mom_3)B^{-6/6}
,\spec\Bigl(\frac{1}{B}\Ham_m(B)\Bigr)\biggr] \leq\widehat{\k}_{0} B^{-7/6}.
\end{equation}
\end{itemize}
\end{thm}
We emphasize that the constants $\lambda_j$, $j=0,\ldots,5$ and $\hmom_0$, 
$\hmom_1$ and $\hmom_2$ agree with the constants in Theorem~\ref{thm:main}. In 
particular $\lambda_0$, $\lambda_1$ and $\lambda_2$ are given 
in~\eqref{eq:firstlambdas}.

We will spend the rest of this section to prove this theorem. We start by 
proving part (iii), which is most detailed, and then go back and implement the
necessary modifications for~(i) and~(ii). 

\subsection{Proof of Theorem~\ref{thm:goodtrial} (iii)}

\subsubsection{Outline}
We will use a trial function $\widehat{\psi}$ on the support of which the 
operators $\hQ_m(B)$ and $\Ham_m(B)$ agree (after a change of coordinates), see
Section~\ref{sec:endgoodtrial} below. This implies that it is enough to prove 
the theorem for $\hQ_m(B)$ instead of $\Ham_m(B)$.

We start by expanding the operator $\hQ_m(B)$ in powers of $B^{-1/6}$. Then we
use the Gru{\v{s}}in method~\cite{gr,sj} to produce a trial state that agrees 
with~\eqref{eq:goodtrial}. 

\subsubsection{Expansion of $\hQ_m(B)$}
We will write
\begin{equation*}
\mom_3= m-\frac{B}{2}-\mom_0\sqrt{B}-\mom_1 B^{1/3}-\mom_2 B^{1/6},
\end{equation*}
and once the optimal values of $\mom_0$, $\mom_1$ and $\mom_2$ are determined we
will write those as $\hmom_0$, $\hmom_1$ and $\hmom_2$ respectively and insert
them in the definition of $\mom_3$.

We expand the operator $\h:=\hQ_m(B)$ in the form
\begin{equation}\label{eq:hexpansion}
 \h \sim \sum_{j= 0}^\infty \h_j B^{-j/6}.
\end{equation}
The expansion~\eqref{eq:hexpansion} is to be understood as follows: For any 
function $f$ in $\mathcal{S}(\mathbb{R}^2_+)$ and for all $N$ it holds that
\begin{equation*}
\h f = \sum_{j=0}^N B^{-j/6} \h_j f  + \bigo\bigl(B^{-(N+1)/6}\bigr)
\end{equation*}
in the sense of $L^2(\mathbb{R}^2_+)$. We recall the formula~\eqref{eq:hQm} for
$\hQ_m(B)$ and expand each term in a Taylor series, valid for small values of 
$B^{-1/3}\rho$ and $B^{-1/2}\tau$. We will need the first seven operators,
\begin{equation}\label{eq:sevenop}
\begin{aligned}
 \h_0 &= -\PD{}{\tau}^2+(\tau+\mom_0)^2,\\
 \h_1 &= 2\Bigl(\mom_1+\frac{\rho^2}{2}\Bigr)(\tau+\mom_0),\\
 \h_2 &= -\PD{}{\rho}^2+\Bigl(\mom_1+\frac{\rho^2}{2}\Bigr)^2 +
2\mom_2(\tau+\mom_0),\\
 \h_3 &= 2\mom_3(\tau+\mom_0) +2\mom_2\Bigl(\mom_1+\frac{\rho^2}{2}\Bigr)
+\tilde{\h}_3,\\
 \h_4 &= 2\mom_3\Bigl(\mom_1+\frac{\rho^2}{2}\Bigr)+\mom_2^2+\tilde{\h}_4,\\
 \h_5 &= 2\mom_2\mom_3 +\tau(3\tau+4\mom_0)\mom_2+\tilde{\h}_5,\\
 \h_6 &= \mom_3^2 + \tau(3\tau+4\mom_0)\mom_3+2\bigl((\mom_0+\tau)\rho^2
+2\tau\mom_1\bigr)\mom_3+\tilde{\h}_6.
\end{aligned}
\end{equation}
with
\begin{align*}
\tilde{\h}_3 &=2\PD{}{\tau} +\tau(\tau+2\mom_0)(\tau+\mom_0)\\
\tilde{\h}_4 &=\mom_0^2\rho^2
+\frac{1}{2}\bigl(6\mom_1+\rho^2\bigr)\tau^2
+4\mom_0\Bigl(\mom_1+\frac{\rho^2}{2}\Bigr)\tau\\
\tilde{\h}_5 &=-2\tau\PD{}{\rho}^2+\frac{\rho^4}{6}(4\mom_0+\tau)
+2\mom_1\rho^2(\tau+\mom_0)+2\mom_1^2\tau\\
\tilde{\h}_6 &=\rho\PD{}{\rho} + 2\tau\PD{}{\tau}
+\frac{1}{12}\rho^6+\frac{2\mom_1}{3}\rho^4
+\mom_1^2\rho^2+\frac{5}{4}\tau^4+4\mom_0\tau^3+3\mom_0^2\tau^2
\end{align*}
\begin{remark}\label{rem:tildeop}
The operators $\tilde{\h}_j$, $3\leq j\leq 6$ are chosen to be independent of 
$\mom_2$ and $\mom_3$. For future reference, we also note that the linear terms 
in $\mom_2$ in the operators $\h_3$, $\h_4$ and $\h_5$ re-appear as linear terms
in $\mom_3$ in the operators $\h_4$, $\h_5$ and $\h_6$ respectively. 
\end{remark}
We study asymptotic expansions on the form
\begin{equation*}
 \lambda \sim \sum_{j= 0}^\infty\lambda_j B^{-j/6}
\quad \text{and}\quad
 \psi(\tau,\rho)\sim\sum_{j= 0}^\infty \psi_j(\tau,\rho)B^{-j/6}.
\end{equation*}
We want to find $\lambda_j$ (as small as possible!) and $\psi_j$ such that
\begin{equation}
\label{eq:simeq}
 (\h-\lambda)\psi \sim 0.
\end{equation}
The expansion~\eqref{eq:simeq} is to be understood term-wise, i.e.,
\begin{equation}\label{eq:firststepeq}
\sum_{j=0}^k (\h_j-\lambda_j)\psi_{k-j}=0.
\end{equation}

To start the Gru{\v{s}}in approach, we need a function to project on. A 
study of~\eqref{eq:firststepeq} for $k=0$ provides us with that.

\subsubsection{Order $B^{0/6}$, a starting point}
At this order equation~\eqref{eq:firststepeq} reads
\begin{equation*}
 (\h_0-\lambda_0)\psi_0 = 0.
\end{equation*}
Notice that $\h_0$ does not act in the $\rho$ variable.
We let $\psi_0(\tau,\rho)=u_0(\tau)\phi_0(\rho)$. We do not want $\phi_0$ to be
identically equal to zero, so we are led to solve
\begin{equation*}
-u_0''+(\tau+\mom_0)^2 u_0=\lambda_0 u_0,\quad u_0'(0)=0.
\end{equation*}
This is the eigenvalue equation for the well-known de~Gennes operator 
$\dG(\mom_0)$ from Appendix~\ref{sec:dG}. 
The smallest eigenvalue $\lambda_0$ is simple and given by
\begin{equation*}
\lambda_0=\Theta_0
\end{equation*}
which is obtained for 
\begin{equation}\label{eq:hmomzero}
\mom_0=\hmom_0=\xi_0,
\end{equation}
see Lemma~\ref{lem:thetazero}. The eigenfunction $u_0$ is positive and belongs 
to $\mathcal{S}(\mathbb{R}_+)$ (see Lemma~\ref{lem:degennesschwartz}). We may 
also assume that $u_0$ is normalized, $\int_0^\infty u_0^2\ed\tau = 1$. 

\subsubsection{Higher orders, the Gru{\v{s}}in approach}\label{sec:grusin}

In this subsection we will implement the Gru{\v{s}}in method~\cite{gr,sj}
which provides us with a systematic way of calculating a trial state for 
$\h:=\hQ_m(B)$. We start by introducing some notation. First, we let
\begin{equation*}
\dh = \h-(\h_0-\lambda_0)\sim\sum_{j=1}^\infty \h_j B^{-j/6}+\lambda_0,
\end{equation*}
and notice that
\begin{equation*}
\dh-\lambda \sim \sum_{j=1}^\infty (\h_j-\lambda_j)B^{-j/6}.
\end{equation*}
Next, we introduce operators 
$\Rp\colon L^2(\mathbb{R})\to L^2(\mathbb{R}^2_+)$, 
$\Rm\colon L^2(\mathbb{R}^2_+)\to L^2(\mathbb{R})$ and 
$\E\colon L^2(\mathbb{R}^2_+)\to L^2(\mathbb{R}^2_+)$ (and in the corresponding
Schwartz spaces) as
\begin{equation*}
\begin{aligned}
(\Rp\phi)(\tau,\rho) &=  \phi(\rho)u_0(\tau),\\
(\Rm f)(\rho) &= \int_0^\infty f(\tau,\rho) u_0(\tau)\ed\tau,\quad \text{and}\\
\E &= I\otimes \Rreg.
\end{aligned}
\end{equation*}
Here $\Rreg$ is the regularized resolvent introduced in~\eqref{eq:rreg}.
We introduce the matrix operators $\HH$ and $\Ezero$, both acting in the 
Hilbert space $L^2(\mathbb{R}^2_+)\oplus L^2(\mathbb{R})$, as
\begin{equation*}
\HH =
\begin{pmatrix}
\h-\lambda & \Rp\\
\Rm & 0
\end{pmatrix}
,\quad \text{and}\quad
\Ezero =
\begin{pmatrix}
\E & \Rp\\
\Rm & 0
\end{pmatrix}.
\end{equation*}
By noting that
\begin{equation*}
\begin{pmatrix}
\h_0-\lambda_0 & \Rp\\
\Rm & 0
\end{pmatrix}
\Ezero = I,
\end{equation*}
we see that
\begin{equation}\label{eq:K}
\K:=\HH\Ezero-I = 
\begin{pmatrix}
(\dh-\lambda)\E & (\dh-\lambda)\Rp\\
0 & 0
\end{pmatrix}.
\end{equation}
Notice that $\K=\bigo(B^{-1/6})$ as an operator on the Schwartz space.

Let $N\in\mathbb{N}$. Then by~\eqref{eq:K},
\begin{equation}
\label{eq:back}
\HH\Ezero\sum_{j=0}^N (-1)^j\K^j =I+(-1)^N\K^{N+1} 
=  I+\bigo\bigl(B^{-(N+1)/6}\bigr),\quad \text{as }B\to\infty.
\end{equation}
By~\eqref{eq:K} we see that $\K^j$ is given by
\begin{equation*}
\K^j=
\begin{pmatrix}
[(\dh-\lambda)\E]^j & [(\dh-\lambda)\E]^{j-1}(\dh-\lambda)\Rp\\
0 & 0
\end{pmatrix}.
\end{equation*}
Let us define $\Einf$, $\Einfp$, $\Einfm$ and $\Einfpm$ via
\begin{equation*}
\begin{pmatrix}
\Einf & \Einfp\\
\Einfm & \Einfpm
\end{pmatrix}
:=\Ezero \sum_{j=0}^N (-1)^j\K^j.
\end{equation*}
Then, by~\eqref{eq:back},
\begin{equation}\label{eq:matris}
\begin{aligned}
(\h-\lambda)\Einf+\Rp\Einfm &= 1+\bigo\bigl(B^{-(N+1)/6}\bigr),\\
(\h-\lambda)\Einfp+\Rp\Einfpm &= \bigo\bigl(B^{-(N+1)/6}\bigr),\\
\Rm\Einf &= \bigo\bigl(B^{-(N+1)/6}\bigr),\\
\Rm\Einfp &= 1+\bigo\bigl(B^{-(N+1)/6}\bigr),
\end{aligned}
\end{equation}
 as $B\to\infty$. Assume that
\begin{equation}\label{eq:phiexpansion}
\phi(\rho) = \sum_{j=0}^{N} \phi_j(\rho)B^{-j/6},\quad \text{with}
\quad \Einfpm\phi = \bigo\bigl(B^{-(N+1)/6}\bigr),\quad \text{as }B\to\infty,
\end{equation}
with $\phi_j\in\mathcal{S}(\mathbb{R})$ for all $j$. Then by inserting the 
vector $(0,\phi)$ in~\eqref{eq:back} and using the second formula 
of~\eqref{eq:matris} we find that
\begin{equation}\label{eq:matristwo}
(\h-\lambda)\Einfp\phi = \bigo\bigl(B^{-(N+1)/6}\bigr),
\quad \text{as }B\to\infty.
\end{equation}
We expand
\begin{equation}\label{eq:EpEpm}
\begin{aligned}
\Einfp &= \sum_{j=0}^N \Ep_j B^{-j/6} + \bigo(B^{-(N+1)/6}),\quad \text{and}\\
\Einfpm & = \sum_{j=0}^{N} \Epm_j B^{-j/6}+ \bigo(B^{-(N+1)/6}),
\quad \text{as }B\to\infty,
\end{aligned}
\end{equation}
where the operators
\begin{equation}
\label{eq:Epj}
(\Ep_jf)(\tau,\rho) =
\mkern-28mu
\sum_{\substack{l_1,\ldots,l_{i}\in\{1,\ldots,j\}\\l_1+\ldots+l_{i}=j}}
\mkern-28mu
(-1)^{i}\Bigl(\prod_{m=1}^i
\E(\h_{l_m}-\lambda_{l_m})\Bigr)u_0(\tau) f(\rho)
\end{equation}
and
\begin{multline}
\label{eq:Epmj}
(\Epm_jf)(\rho) = 
\int_0^\infty u_0(\tau)\Bigg[\sum_{i=0}^{j-1}(-1)^{i-1}\mkern-38mu
\sum_{\substack{l_1,\ldots,l_{i+1}\in\{1,\ldots,j\}\\l_1+\ldots+l_{i+1}=j}}
\mkern-38mu
(\h_{l_1}-\lambda_{l_1})\E (\h_{l_2}-\lambda_{l_2})\E \times \cdots\\ 
\times \E(\h_{l_{i+1}}-\lambda_{l_{i+1}})\Bigg]u_0(\tau) f(\rho) \ed \tau
\end{multline}
are independent of $N$. Here we use the convention that the summand is just 
$(\h_j-\lambda_j)$ for $i=0$. We write
\begin{equation}\label{eq:psijdefeq}
\Einfp\phi =\mkern-14mu\sum_{\substack{0\leq j,k\leq N\\j+k\leq N}}\mkern-14mu 
\Ep_j \phi_{k} B^{-(j+k)/6}+\bigo(B^{-(N+1)/6}),
\quad \text{as }B\to\infty,
\end{equation}
and define for $j\in\{0,\ldots,N\}$ the function $\psi_j$ to be the 
coefficient in front of $B^{-j/6}$ in the sum on the right-hand side
in~\eqref{eq:psijdefeq}. We use~\eqref{eq:Epj} to get a formula for $\psi_j$,
\begin{equation}\label{eq:psijcomplete}
\psi_j(\tau,\rho) = 
\mkern-24mu
\sum_{\substack{l_1,\ldots,l_{i}\in\{1,\ldots,j\}\\l_1+\ldots+l_i+k=j}}
\mkern-24mu(-1)^i 
\Bigl(\prod_{m=1}^i \E (\h_{l_m}-\lambda_{l_m}) \Bigr)u_0 \phi_k.
\end{equation}
We define our trial state by
\begin{equation}\label{eq:psij}
\psi(\tau,\rho) = \sum_{j=0}^{N} \psi_j(\tau,\rho)B^{-j/6}.
\end{equation}
Since the operators involved are continuous (uniformly in $B>1$), considered 
on Schwartz functions (the functions $\psi_j$ will be Schwartz functions, see 
Section~\ref{sec:seveneight}), it follows from~\eqref{eq:matristwo} that
\begin{equation*}
(\h-\lambda)\psi = \bigo(B^{-(N+1)/6}),\quad \text{as }B\to\infty.
\end{equation*}
In particular~\eqref{eq:firststepeq} holds for all $0\leq k\leq N$.

Before we start with the calculations, let us note that the 
condition~\eqref{eq:phiexpansion} on the right reads that
\begin{equation}\label{eq:theeqk}
\sum_{j=1}^k \Epm_j \phi_{k-j}=0,\quad \text{for all }0\leq k\leq N.
\end{equation}
We also introduce the notation
\begin{equation*}
\tEpm_j = \Epm_j-\lambda^{\vphantom{\pm}}_j.
\end{equation*}

\subsubsection{Order $B^{-1/6}$, calculation of $\lambda_1$}
To calculate $\lambda_1$ we need the operator $\Epm_1$. By~\eqref{eq:star} we 
see that, for any function $f\in L^2(\mathbb{R})$, it holds that
\begin{equation*}
\Epm_1f = -\Rm(\h_1-\lambda_1)\Rp f =
-\Rm\biggl(2\Bigl(\mom_1+\frac{\rho^2}{2}\Bigr)(\tau+\hmom_0) -
\lambda_1\biggr)\Rp f=\lambda_1 f,
\end{equation*}
and so $\Epm_1$ is just a multiplication operator. The 
equation~\eqref{eq:theeqk} 
reads for $k=1$
\begin{equation*}
0=\Epm_1\phi_0 =\lambda_1\phi_0.
\end{equation*}
Since we do not want $\phi_0$ to be identically zero, we find that 
\begin{equation*}
\lambda_1=0\quad \text{and}\quad \Epm_1 = 0
\end{equation*}
as an operator. 

\subsubsection{Order $B^{-1/3}$, calculation of $\lambda_2$}
Using that $\Epm_1=0$, the equation~\eqref{eq:theeqk} for $k=2$ reads
\begin{align}
0&=-\Epm_2\phi_0 = -\Rm[\h_1\E\h_1 -(\h_2-\lambda_2)]\Rp\phi_0\nonumber\\
 &=-\int_0^\infty
u_0\biggl[2\Bigl(\mom_1+\frac{\rho^2}{2}\Bigr)(\tau+\hmom_0) \Rreg
2\Bigl(\mom_1+\frac{\rho^2}{2}\Bigr)(\tau+\hmom_0) \nonumber\\
&\qquad -\Bigl(-\dtworho{}+\Bigl(\mom_1+\frac{\rho^2}{2}\Bigr)^2 -
\lambda_2\Bigr)\biggr]\phi_0 u_0\ed\tau\nonumber\\
 &= \biggl(-\dtworho{}+(1-4k_1)\Bigl(\mom_1+\frac{\rho^2}{2}\Bigr)^2 -
\lambda_2\biggr)\phi_0.\label{eq:montgomery}
\end{align}
The number $k_1$ is defined in~\eqref{eq:kj} and satisfies $(1-4k_1)=\dGdiff$ 
by~\eqref{eq:respar}. We choose $\lambda_2=\lambda_2(\mom_1)$ to be the
smallest eigenvalue of the operator 
$-\dtworho{}+\dGdiff\bigl(\mom_1+\frac{\rho^2}{2}\bigr)^2$, 
and optimize in $\mom_1$,
using the analysis in Appendix~\ref{sec:Mg}, to get
\begin{equation}\label{eq:hmomone}
\lambda_2 = 2^{-2/3}\hat{\nu}_0\dGdiff^{1/3}=\widehat{\gamma}_0
,\quad \text{and}\quad 
\mom_1 = \hmom_1 = \tilde{\mom} = (2\dGdiff)^{-1/3}\widehat{\mom}.
\end{equation}
From now on $\phi_0$ denotes the corresponding eigenfunction, normalized
in $L^2(\mathbb{R})$. We note
that this also fixes the function $\psi_0(\tau,\rho)=u_0(\tau)\phi_0(\rho)$.

\subsubsection{Order $B^{-k/6}$, $k\geq 3$}
On the level $B^{-k/6}$, our unknowns are $\lambda_k$, $\phi_{k-2}$ and also, up
to some level, $\mom_2$ and $\mom_3$. The following procedure will determine
$\phi_{k-2}$ and $\lambda_k$. For $\phi_{k-2}$ to exist, it must be possible to
solve
\begin{equation*}
-\Epm_2\phi_{k-2}=\sum_{j=3}^k \Epm_j\phi_{k-j}.
\end{equation*}
As we saw above $-\Epm_2$ is the Montgomery operator minus its 
lowest eigenvalue. Thus, we want the solvability of the differential equation
\begin{equation}
\label{eq:rhosolveeq}
\left(-\dtworho{}+\dGdiff\Bigl(\hmom_1+\frac{\rho^2}{2}\Bigr)^2 -
\lambda_2\right)\phi_{k-2}=\sum_{j=3}^k \Epm_j\phi_{k-j}.
\end{equation}
But this is equivalent to the condition that the right hand side is orthogonal
to the ground state $\phi_0$, i.e.
\begin{equation}
\label{eq:orth}
\int_{-\infty}^\infty \phi_0\biggl(\sum_{j=3}^k \Epm_j\phi_{k-j}\biggr)\ed\rho
= 0.
\end{equation}
From~\eqref{eq:orth} we get a formula for the unknown $\lambda_k$,
\begin{equation*}
\lambda_k = -\int_{-\infty}^\infty \phi_0\biggl(\sum_{j=3}^{k-1}
\Epm_j\phi_{k-j}+\tEpm_k\phi_0\biggr)\ed\rho.
\end{equation*}
Note that the right hand side will at some levels depend on $\mom_2$ and
$\mom_3$. As soon as it does, we will minimize $\lambda_k$
over $\mom_2$ and $\mom_3$. When $\lambda_k$ is determined we
find $\phi_{k-2}$ by inverting the Montgomery operator in~\eqref{eq:rhosolveeq}.

We calculate the first terms, the coefficients that appear are introduced in 
Appendix~\ref{sec:modeloperators}.

\subsubsection{Order $B^{-1/2}$, calculation of $\lambda_3$}

For $\lambda_3$ we get
\begin{equation*}
\begin{aligned}
\lambda_3 &= -\int_{-\infty}^\infty \phi_0\bigl(\tEpm_3\phi_0\big)\ed\rho\\
          &= \big\langle \phi_0 u_0,\big[\h_3
              - \h_1\E(\h_2-\lambda_2) - (\h_2-\lambda_2)\E\h_1
              + \h_1\E\h_1\E\h_1\big]u_0\phi_0\big\rangle.
\end{aligned}
\end{equation*}
Calculations, using the formulas~\eqref{eq:sevenop} for $\h_j$ and the 
calculations in Appendix~\ref{sec:modeloperators} give
\begin{equation*}
\begin{aligned}
\langle \phi_0 u_0, \h_3 u_0\phi_0 \rangle &= 
-\frac{7}{6}u_0(0)^2 +\hmom_0^3-\frac{1}{2}\hmom_0^2,\\
\langle \phi_0 u_0,\h_1\E(\h_2-\lambda_2)u_0\phi_0\rangle &=0,\\
\langle \phi_0 u_0 ,(\h_2-\lambda_2)\E\h_1u_0\phi_0\rangle &=0,\\
\langle \phi_0 u_0 ,\h_1\E\h_1\E\h_1 u_0\phi_0 \rangle & = 8k_2M_{0,0}^3,
\end{aligned}
\end{equation*}
which implies that
\begin{equation}
\label{eq:lthree}
\lambda_3 = -\frac{7}{6}u_0(0)^2+\hmom_0^3-\frac{1}{2}\hmom_0^2 + 8k_2M_{0,0}^3.
\end{equation}

\subsubsection{Order $B^{-2/3}$, calculation of $\lambda_4$}
\label{sec:lfourcalc}
For $\lambda_4$ we get
\begin{equation*}
\begin{aligned}
\lambda_4 &= -\int_{-\infty}^\infty \phi_0
              \bigl(\Epm_3\phi_1+\tEpm_4\phi_0\bigr)\ed\rho\\
           &=  \big\langle \phi_0 u_0,\big[(\h_3-\lambda_3) 
            - \h_1\E(\h_2-\lambda_2)-(\h_2-\lambda_2)\E\h_1
            +\h_1\E\h_1\E\h_1\big]u_0 \phi_1\big\rangle\\
           &\quad +\big\langle \phi_0 u_0, \big[\h_4
            -\h_1\E(\h_3-\lambda_3) -(\h_2-\lambda_2)\E(\h_2-\lambda_2)
                   -(\h_3-\lambda_3)\E\h_1\\
           & \qquad +\h_1\E\h_1\E(\h_2-\lambda_2)+\h_1\E(\h_2-\lambda_2)\E\h_1
                   +(\h_2-\lambda_2)\E\h_1\E\h_1\\
           & \qquad -\h_1\E\h_1\E\h_1\E\h_1\big]u_0\phi_0 \big\rangle.\\
\end{aligned}
\end{equation*}
Instead of calculating all these integrals explicitly we only look for their
dependence on $\mom_2$ and $\mom_3$. It turns out that they are all independent 
of $\mom_3$ and that $\mom_2$ appears in the following integrals, which we
calculate with help of the relations in Appendix~\ref{sec:modeloperators} (we 
let $\text{const}$ denote any constant independent of $\mom_2$):
The quadratic dependence on $\mom_2$ is given by
\begin{align*}
\langle \phi_0 u_0 ,\h_4 u_0\phi_0 \rangle
&= \mom_2^2 +\text{const},\\
-\langle \phi_0 u_0 ,(\h_2-\lambda_2) \E (\h_2-\lambda_2) u_0\phi_0 
\rangle &= -4k_1\mom_2^2.
\end{align*}
The linear dependence on $\mom_2$ is given by
\begin{equation}\label{eq:momtwolin}
\begin{aligned}
\langle \phi_0 u_0 ,(\h_3-\lambda_3) u_0\phi_1 \rangle 
&= 2\mom_2 M_{0,1}^1+\text{const},\\
-\langle \phi_0 u_0 ,\h_1\E(\h_2-\lambda_2) u_0\phi_1 \rangle
& = -4 M_{0,1}^1 k_1 \mom_2,\\
-\langle \phi_0 u_0, (\h_2-\lambda_2)\E\h_1 u_0\phi_1 \rangle
&= -4 M_{0,1}^1 k_1 \mom_2,\\
\langle \phi_0 u_0 ,\h_1\E\h_1\E(\h_2-\lambda_2) u_0\phi_0 \rangle
 &= 8M_{0,0}^2k_2 \mom_2,\\
\langle \phi_0 u_0, \h_1\E(\h_2-\lambda_2)\E\h_1 u_0\phi_0 \rangle
 &= 8M_{0,0}^2k_2 \mom_2 + \text{const},\\
\langle \phi_0 u_0, (\h_2-\lambda_2)\E\h_1\E\h_1 u_0\phi_0 \rangle
&= 8M_{0,0}^2k_2 \mom_2.
\end{aligned}
\end{equation}
The result is
\begin{equation}
\label{eq:lfour}
\begin{aligned}
\lambda_4 & = \bigl(1-4k_1\bigr)\mom_2^2 
          +2\bigl(M_{0,1}^1(1-4k_1)+12M_{0,0}^2k_2\bigr)\mom_2 +\text{const}\\
          & = \dGdiff\mom_2^2 
          +2\bigl(M_{0,1}^1\dGdiff+12M_{0,0}^2k_2\bigr)\mom_2 +\text{const}.\\
\end{aligned}
\end{equation}
We see that $\lambda_4$ depends on $\mom_2$ as a parabola and is minimal if
\begin{equation}
\label{eq:dtwo}
\mom_2 =\hmom_2 = -\frac{1}{\dGdiff}\left(M^1_{0,1}\dGdiff+ 
12M^2_{0,0}k_2\right)=-M^1_{0,1}-M^2_{0,0}12k_2\dGdiff^{-1}.
\end{equation}
This fixes the value of $\lambda_4$.

\subsubsection{Order $B^{-5/6}$, calculation of $\lambda_5$}
\label{sec:lfivecalc}
For $\lambda_5$ we get
\begin{equation*}
\lambda_5 = -\int_{-\infty}^\infty \phi_0\bigl(\Epm_3\phi_2+\Epm_4\phi_1+
                                              \tEpm_5\phi_0\bigr)\ed\rho.
\end{equation*}
To calculate all the integrals corresponding to $\lambda_5$ in full detail 
would be too cumbersome for the presentation. 
The importance of this step is the dependence on $\mom_3$. 
Thus we only calculate the integrals involving $\mom_3$.
First, we get some integrals involving $\mom_3$ and $\hmom_2$.
\begin{align*}
\langle \phi_0 u_0 , \h_5   u_0\phi_0 \rangle 
&= 2\hmom_2 \mom_3,\\
-\langle \phi_0 u_0 , (\h_2 - \lambda_2)  
  \E   (\h_3 - \lambda_3)   u_0\phi_0 \rangle 
&= -4k_1\hmom_2\mom_3+\text{const},\\
-\langle \phi_0 u_0  (\h_3 - \lambda_3)  
  \E   (\h_2 - \lambda_2)   u_0\phi_0 \rangle 
&= -4k_1\hmom_2\mom_3+\text{const}.\\
\end{align*}
Here $\text{const}$ denotes a constant that is independent of $\mom_3$.
By Remark~\ref{rem:tildeop} the linear terms in $\mom_3$ (independent of 
$\hmom_2$) are the same as the linear terms in $\mom_2$ for $\lambda_4$, 
see~\eqref{eq:momtwolin}.
The result is that $\lambda_5$ is given by
\begin{equation*}
\lambda_5 = \bigl[2M_{0,1}^1(1-4k_1)+2(1-4k_1)\hmom_2+24M_{0,0}^2k_2\bigr]\mom_3 
+ \text{const}.
\end{equation*}
By the choice of $\hmom_2$ in~\eqref{eq:dtwo} we see that the coefficient in 
front of $\mom_3$ is zero, so $\lambda_5$ is independent of $\mom_3$. We 
continue with the calculation of the dependence of $\mom_3$ in $\lambda_6$.

\subsubsection{Order $B^{-1}$, calculation of $\lambda_6$}

We do not calculate $\lambda_6$ in its full detail. The important part is its 
dependence on $\mom_3$. There are two integrals that give rise to quadratic 
terms $\mom_3^2$:
\begin{align*}
\langle \phi_0 u_0 , \h_6   u_0\phi_0\rangle 
&= \mom_3^2+\text{lower order terms in $\mom_3$},\quad \text{and,}\\
-\langle \phi_0 u_0,  (\h_3-\lambda_3)\E(\h_3-\lambda_3) u_0\phi_0 \rangle
&= -4k_1 \mom_3^2+\text{lower order terms in $\mom_3$}.
\end{align*}
The result is that $\lambda_6$ depends on $\mom_3$ quadratically
with coefficient $1-4k_1=\dGdiff>0$ in front of $\mom_3^2$. We introduce the 
constants $\hmom_3$ and $\C$ via
\begin{equation}\label{eq:lsix}
\lambda_6=\lambda_6(\mom_3) = \dGdiff (\mom_3-\widehat{\mom}_3)^2+\C.
\end{equation}

\subsubsection{Order $B^{-7/6}$ and $B^{-8/6}$, regularity properties}
\label{sec:seveneight}
We note in order to obtain the functions $\phi_5$ and $\phi_6$ we should 
continue the calculations to the scales $B^{-7/6}$ and $B^{-8/6}$ respectively.

We end this section by noting that the facts that $u_0$ and $\phi_0$ are 
Schwartz functions (Lemmas~\ref{lem:degennesschwartz} 
and~\ref{lem:montgomeryschwartz}) 
and that the resolvents of the de~Gennes and Montgomery operators both 
maps the corresponding Schwartz space continuously to itself 
(Lemmas~\ref{lem:degennesschwartz} and~\ref{lem:rhoexpdec}), 
imply that the functions $\psi_j$, $j=0,\ldots,6$, given 
by~\eqref{eq:psijcomplete} all belong to $\mathcal{S}(\mathbb{R}^2_+)$.

\subsubsection{End of proof of Theorem~\ref{thm:goodtrial} (iii)}
\label{sec:endgoodtrial}
The calculations above provide us with $\lambda_j$, and 
functions $\psi_j\in \mathcal{S}(\mathbb{R}^2_+)$, $j=0,\ldots,6$.
We note that among the constants $\lambda_j$, $\lambda_6$ is the only one that
depends on $\mom_3$, as in~\eqref{eq:lsix}. Moreover, by carefully following
$\mom_3$ through the calculations of $\psi_j$, we find that $\psi_0$, $\psi_1$
and $\psi_2$ do not depend on $\mom_3$, while
\begin{equation*}
\begin{aligned}
\psi_3 & =\psi_{3,0}+\psi_{3,1}\mom_3\\
\psi_4 & =\psi_{4,0}+\psi_{4,1}\mom_3+\psi_{4,2}\mom_3^2\\
\psi_5 & =\psi_{5,0}+\psi_{5,1}\mom_3+\psi_{5,2}\mom_3^2+\psi_{5,3}\mom_3^3\\
\psi_6 & =\psi_{6,0}+\psi_{6,1}\mom_3+\psi_{6,2}\mom_3^2+\psi_{6,3}\mom_3^3
           + \psi_{6,4}\mom_3^4,\\
\end{aligned}
\end{equation*}
where all the involved functions belong to $\mathcal{S}(\mathbb{R}^2_+)$.
We let $\chi_B$ be a usual smooth cut-off function, satisfying 
\begin{align}\label{eq:cutofffunc}
\chi_B(\tau,\rho)=1\ \text{on}\ &\Bigl\{(\tau,\rho)
\mid 0<\tau<\frac16 B^{1/2},\ 
|\rho|<\frac{\pi}{8}B^{1/3}\Bigr\},\quad\text{and}\\
\supp\bigl(\chi_B(\tau,\rho))\subset&\Bigl\{(\tau,\rho)
\mid 0<\tau<\frac13 B^{1/2},\ 
|\rho|<\frac{\pi}{4}B^{1/3}\Bigr\},\nonumber
\end{align}
and with Neumann condition at $\tau=0$, $(\PD{\chi_B}{\tau})(0,\rho)=0$.
With
\begin{equation*}
\lambda = \sum_{j=0}^6 \lambda_j B^{-j/6}
,\quad
\hpsi(\tau,\rho)=\chi_B\sum_{j=0}^6 \psi_jB^{-j/6}
,\quad\text{and}\quad \hQ_m = \sum_{j=0}^6 \h_j B^{-j/6}+R_7
\end{equation*}
we can write
\begin{multline*}
(\hQ_m-\lambda)\hpsi = \sum_{k=0}^6 \sum_{j=0}^k
(\h_j-\lambda_j)\psi_{k-j} B^{-k/6}
+ \sum_{k=1}^{6} \sum_{j=k}^6 (\h_j-\lambda_j)\psi_{k+6-j} B^{-(k+6)/6}
+ R_7 \hpsi\\
+\sum_{k=0}^6 \sum_{j=0}^k(\h_j-\lambda_j)(1-\chi_B)\psi_{k-j} B^{-k/6}
+ \sum_{k=1}^{6} \sum_{j=k}^6 
(\h_j-\lambda_j)(1-\chi_B)\psi_{k+6-j} B^{-(k+6)/6}.
\end{multline*}
The first sum vanishes according to~\eqref{eq:firststepeq}. Since 
$|\mom_3|\leq \widetilde{\k}_0$ the second sum and 
$R_7\hpsi$ are both bounded by a constant (independent of $\mom_3$) times 
$B^{-7/6}$. The last two sums are of order $\bigo(B^{-\infty})$, since all 
functions $\psi_j\in\mathcal{S}(\mathbb{R}^{2}_+)$. We therefore get the 
existence of constants $\widehat{\k}_0$ and $\widehat{B}_0$ such that
\begin{equation*}
\bigl\|(\hQ_m-\lambda)\hpsi\bigr\|\leq \widehat{\k}_0 B^{-7/6}\bigl\|\hpsi\bigr\|
\end{equation*}
for $B>\widehat{B}_0$. By the spectral theorem we conclude~\eqref{eq:goodtrial}.
\qed

\subsection{Proof of Theorem~\ref{thm:goodtrial} (i)}
We use the same approach as in the proof of (iii). Actually, we
repeat the same calculations, but with the differences that $\mom_3=0$ and 
$\mom_2=\hmom_2+\para$. The result is that the operators $\h_0$ and $\h_1$ are 
independent of $\para$, while $\h_2$ involves $\para$ linearly in the form 
$2\para(\tau+\mom_0)$, $\h_3$ involves $\para$ linearly and $\h_4$ 
quadratically. By keeping track of the $\para$ in the calculations, we find that
$\lambda$ will have the form
\begin{equation}\label{eq:lambdafour}
\lambda = \lambda_0 + \lambda_2 B^{-2/6}+\lambda_3 B^{-3/6} 
+\lambda_4(\para)B^{-4/6},
\end{equation}
where $\lambda_0=\Theta_0$, $\lambda_2=\widehat{\gamma}_0$, 
$\lambda_3$ is given in~\eqref{eq:lthree} and 
$\lambda_4(\para)=\lambda_4+\dGdiff \para^2$ where $\lambda_4$ is the 
coefficient calculated in Section~\ref{sec:lfourcalc} and $\dGdiff>0$ is the 
constant from Lemma~\ref{lem:thetazero}.

The trial state $\hpsi$ has the form
\begin{multline}\label{eq:psifour}
\hpsi=\chi_B\Bigr(\psi_0+\psi_1 B^{-1/6}+ (\psi_{2,0}+\psi_{2,1}\para)B^{-2/6}
+(\psi_{3,0}+\psi_{3,1}\para+\psi_{3,2}\para^2)B^{-3/6}\\
+(\psi_{4,0}+\psi_{4,1}\para+\psi_{4,2}\para^2+\psi_{4,3}\para^3)B^{-4/6}\Bigr),
\end{multline}
where all involved functions belong to $\mathcal{S}(\mathbb{R}^2_+)$. 

We write $\hQ_m(B)=\sum_{j=0}^4 \h_j B^{-j/6}+R_5$
and organize the terms in $(\hQ_m(B)-\lambda)\hpsi$ as
\begin{align*}
(\hQ_m(B)-\lambda)\hpsi=& 
\sum_{k=0}^4\sum_{j=0}^4(\h_j-\lambda_j)\psi_{k-j} B^{-k/6}
+\sum_{k=1}^4\sum_{j=k}^4 (\h_j-\lambda_j)\psi_{k+4-j}B^{-(k+4)/6}\\ 
&+ R_{5}\hpsi
+\sum_{k=0}^4\sum_{j=0}^4(\h_j-\lambda_j)(1-\chi_B)\psi_{k-j} B^{-k/6}\\
&+\sum_{k=1}^4\sum_{j=k}^4 (\h_j-\lambda_j)(1-\chi_B)\psi_{k+4-j}B^{-(k+4)/6}.
\end{align*}
The first double sum is zero by~\eqref{eq:firststepeq}. For the second one, we
use~\eqref{eq:psifour} and~\eqref{eq:lambdafour} to bound it in $L^2$-norm by 
(recall that $|\para|\leq \widetilde{\k}_{2/6} B^{1/6}$)
\begin{equation*}
C'(1+|\para|^2) B^{-5/6}
\bigl(1+|\para|B^{-1/6}+|\para|^2B^{-2/6}+|\para|^3 B^{-3/6}\bigr)
\leq C''(1+|\para|^2)B^{-5/6}.
\end{equation*}
Since the original operator depends quadratically on $\para$, it follows that
\begin{equation*}
\|R_5\hpsi\|\leq (1+|\para|^2)B^{-5/6}\|\hpsi\|.
\end{equation*}
Again, the last two sums are $\bigo(B^{-\infty})$, since the involved functions
belong to the Schwartz space. We get, with $\lambda$ as 
in~\eqref{eq:lambdafour}, that
there exist constants $\widehat{\k}_{2/6}$ and $\widehat{B}_{2/6}$ such that
\begin{equation*}
\bigl\|(\hQ_m-\lambda)\hpsi\bigr\|
\leq \widehat{\k}_{2/6}(1+|\para|^2)B^{-5/6}\bigl\|\hpsi\bigr\|
\end{equation*}
for $B>\widehat{B}_{2/6}$. By the spectral theorem we 
conclude~\eqref{eq:goodfour}.\qed

\subsection{Proof of Theorem~\ref{thm:goodtrial} (ii)}
We repeat the Gru{\v{s}}in calculation from the previous step once more, 
but this time we take one more term in the expansion. The result is a 
function $\hpsi$ of the form
\begin{align*}
\hpsi=\chi_B\Bigl(&\psi_0+\psi_1 B^{-1/6}+ (\psi_{2,0}+\psi_{2,1}\para)B^{-2/6}
+(\psi_{3,0}+\psi_{3,1}\para+\psi_{3,2}\para^2)B^{-3/6}\\
&+(\psi_{4,0}+\psi_{4,1}\para+\psi_{4,2}\para^2+\psi_{4,3}\para^3)B^{-4/6}\\
&+(\psi_{5,0}+\psi_{5,1}\para+\psi_{5,2}\para^2+\psi_{5,3}\para^3
+ \psi_{5,4}\para^4)B^{-5/6}\Bigr)
\end{align*}
where all involved functions belongs to $\mathcal{S}(\mathbb{R}^2_+)$. We also
get 
\begin{equation}\label{eq:lambdafive}
\lambda = \lambda_0 + \lambda_2 B^{-2/6}+\lambda_3 B^{-3/6} 
+\lambda_4(\para)B^{-4/6} +\lambda_5(\para)B^{-5/6},
\end{equation}
with the same $\lambda_0$, $\lambda_2$, $\lambda_3$ and $\lambda_4(\para)$ as
in~\eqref{eq:lambdafour}. Moreover, $\lambda_5(\para)$ depends on $\para$ 
as a polynomial with constant coefficient $\lambda_5$, calculated in 
Section~\ref{sec:lfivecalc},
\begin{equation*}
\lambda_5(\para)=\lambda_5+a_1\para+a_2\para^2+a_3\para^3
\end{equation*}
for some $a_1$, $a_2$ and $a_3$ in $\mathbb{R}$. 

We write $\hQ_m(B)=\sum_{j=0}^5\h_j B^{-j/6}+R_6$ and organize the terms in
$(\hQ_m-\lambda)\hpsi$ as
\begin{align*}
(\hQ_m(B)-\lambda)\hpsi=& 
\sum_{k=0}^5\sum_{j=0}^5 (\h_j-\lambda_j)\psi_{k-j} B^{-k/6}
+\sum_{k=1}^5\sum_{j=k}^5 (\h_j-\lambda_j)\psi_{k+5-j}B^{-(k+5)/6}\\ 
&+ R_{6}\hpsi
+\sum_{k=0}^5\sum_{j=0}^5 (\h_j-\lambda_j)(1-\chi_B)\psi_{k-j} B^{-k/6}\\
&+\sum_{k=1}^5\sum_{j=k}^5 (\h_j-\lambda_j)(1-\chi_B)\psi_{k+5-j}B^{-(k+5)/6}.
\end{align*}
Again, the first double sum is zero. The second two 
terms are of order $B^{-6/6}$, uniformly for bounded $\para$,
and the last two sums are of order $\bigo(B^{-\infty})$.
We get, with $\lambda$ as in~\eqref{eq:lambdafive}, the existence of constants 
$\widehat{\k}_{1/6}$ and $\widehat{B}_{1/6}$ such that
\begin{equation*}
\bigl\|(\hQ_m-\lambda)\hpsi\bigr\|
\leq \widehat{\k}_{1/6}B^{-6/6}\bigl\|\hpsi\bigr\|
\end{equation*}
for $B>\widehat{B}_{1/6}$. We use the spectral theorem to conclude 
inequality~\eqref{eq:goodfive}.
\qed

%
%
%
%
%
%
%  REFINED LOWER BOUNDS
%
%
%
%
%
%
%

\section{Refined lower bounds}\label{sec:newlower}

\subsection{Statement}
We combine the lower bounds from Section~\ref{sec:lower} with part (i) and (ii)
from Theorem~\ref{thm:goodtrial} to have the improved lower bound.

\begin{thm}\label{thm:refinedlower}
Let $K>0$. With the constants $\hmom_0$, $\hmom_1$, $\hmom_2$,  
$\lambda_j$, $j=0,\ldots,5$ and $\dGdiff$, and with $\mom_3$  
from Theorem~\ref{thm:goodtrial} 
there exist constants $B_{0}$ and $\k_{0}$ such that 
if $|\mom_3|\geq \k_{0}$ and $B>B_0$ then
\begin{equation}\label{eq:refinedlower}
\eigone{\Ham_m(B)}\geq \lambda_0B+\lambda_2 B^{4/6}+\lambda_3 B^{3/6}
+\lambda_4 B^{2/6}+\lambda_5 B^{1/6}+K.
\end{equation}
\end{thm}

\subsection{Proof}
We divide the proof into several parts, depending on the size of $\mom_3$. 
Actually, for most values of $\mom_3$ we prove stronger results. We remind the
reader that $\lambda_0=\Theta_0$ and $\lambda_2=\widehat{\gamma}_0$.

\subsubsection{Proof for $\k_{1/2} B^{1/2}\leq |\mom_3|$}
It follows from Lemmas~\ref{lem:psicutoff} and~\ref{lem:firstmombound} that 
there exist constants $\k_{1/2}$ and $B_{1/2}$ such that for all 
$m\in\mathbb{Z}$, $|\mom_3|\geq \k_{1/2}B^{1/2}$, $B>B_{1/2}$ and for all 
$\psi$ that satisfies $\tQ_m(B)\psi=\lambda(B)\psi$ with
$\lambda(B)\leq \Theta_0B+\omega B^{2/3}$ it holds that
\begin{equation*}
\tq_m[\psi]\geq (\Theta_0+1)\|\psi\|^2,
\end{equation*}
which clearly implies~\eqref{eq:refinedlower}.

\subsubsection{Proof for $\k_{1/3} B^{1/3}\leq |\mom_3|\leq \k_{1/2} B^{1/2}$}
Assume that $|\mom_3|\leq \k_{1/2}B^{1/2}$ and that $B>B_{1/2}$. 
By Lemmas~\ref{lem:rholocone} and~\ref{lem:rholoctwo} it follows that there 
exist constants $\k_{1/3}$ and $B_{1/3}$ such that for all $m\in\mathbb{Z}$,
$|\mom_3|\geq \k_{1/3}B^{1/3}$ and $B>B_{1/3}$ it holds that
\begin{equation*}
\q_m[\psi]\geq \bigl(\Theta_0+(\widehat{\gamma}_0+1)B^{-1/3}\bigr)
\|\psi\|^2,\quad 
\forall \psi\in\dom(\q_m).
\end{equation*}
By possibly changing $\k_{1/3}$ and $B_{1/3}$ slightly, it follows from 
Lemmas~\ref{lem:psicutoff} and~\ref{lem:bhalf} that for all $m\in\mathbb{Z}$
such that $|\mom_3|\geq \k_{1/3}B^{1/3}$, $B>B_{1/3}$ and for all $\psi$ that 
satisfies $\tQ_m(B)\psi=\lambda(B)\psi$ with
$\lambda(B)\leq \Theta_0B+\omega B^{2/3}$ it holds that
\begin{equation*}
\tq_m[\psi]\geq \bigl(\Theta_0+(\widehat{\gamma}_0+1)
B^{-1/3}\bigr)\|\psi\|^2,
\end{equation*}
from which~\eqref{eq:refinedlower} follows.

\subsubsection{Proof for $\k_{5/16} B^{5/16}\leq |\mom_3|\leq \k_{1/3} B^{1/3}$}
Assume that $|\mom_3|\leq \k_{1/3}B^{1/3}$ and that $B>B_{1/3}$ with the 
constants $\k_{1/3}$ and $B_{1/3}$ from the previous step. 
We define $\para$ by the equation
\begin{equation*}
m = \frac{B}{2}+\hmom_0 B^{1/2}+(\hmom_1+\para)B^{1/3},
\end{equation*}
where $\hmom_0$ and $\hmom_1$ are the constants from \eqref{eq:hmomzero} 
and~\eqref{eq:hmomone}, and note that the condition $|\mom_3|<\k_{1/3}B^{1/3}$ 
implies that $|\delta|\leq C$ for some constant $C$. 
From Lemma~\ref{lem:montgomery} it follows that for all $|\para|\leq C$ there 
exist a positive constant $C_{\text{pos}}$ such that
\begin{equation}\label{eq:Mgexp}
\eigone{\widetilde{\M}}(\hmom_1+\para)\geq 
\eigone{\widetilde{\M}}(\hmom_1)
+C_{\text{pos}}|\para|^2.
\end{equation}
For $|\para|\geq C_1B^{-1/48}$ we combine~\eqref{eq:Mgexp} with
Proposition~\ref{prop:gap}~(A) to find that
\begin{equation*}
\q_m[\psi]\geq \bigl(\Theta_0 + 
\widehat{\gamma}_0 B^{-1/3} +(C_{\text{pos}}C_1^2-C) B^{-3/8}\bigr)\|\psi\|^2.
\end{equation*}
for sufficiently large $B$. If we choose $C_1$ sufficiently large, we get the
existence of a positive constant $C'_{\text{pos}}$ such that 
for $C_1B^{-1/48}\leq |\para|\leq C$ and for all $B$ large enough it holds that
\begin{equation*}
\q_m[\psi]\geq \bigl(\Theta_0 + 
\widehat{\gamma}_0 B^{-1/3} +C'_{\text{pos}} B^{-3/8}\bigr) 
\|\psi\|^2.
\end{equation*}
Finally, we invoke Lemmas~\ref{lem:psicutoff} and~\ref{lem:bhalf}, decrease
the constant $C'_{\text{pos}}$ slightly to $C''_{\text{pos}}$ if necessary, to 
get existence of positive constants $\k_{5/16}$ and $B_{5/16}$ such that if 
$\k_{5/16}B^{5/16}\leq |\mom_3|\leq \k_{1/3}B^{1/3}$ and $B>B_{5/16}$, then for 
all $\psi$ that satisfies $\tQ_m(B)\psi=\lambda(B)\psi$ with
$\lambda(B)\leq \Theta_0B+\omega B^{2/3}$ it holds that
\begin{equation*}
\tq_m[\psi] \geq \bigl(\Theta_0B +  
\widehat{\gamma}_0 B^{2/3} +C''_{\text{pos}} B^{5/8}\bigr)
\|\psi\|^2.
\end{equation*}
This inequality is also stronger than~\eqref{eq:refinedlower}.

\subsubsection{Proof for $\k_{1/6} B^{1/6}\leq |\mom_3| < \k_{5/16} B^{5/16}$} 
Assume that $|\mom_3| < \k_{5/16} B^{5/16}$ and $B>B_{5/16}$ with the constants 
$\k_{5/16}$ and $B_{5/16}$ from the previous step. We introduce $\para$ as
\begin{equation*}
m = \frac{B}{2}+\hmom_0 B^{1/2}+\hmom_1B^{1/3}+(\hmom_2+\para)B^{1/6},
\end{equation*}
where $\hmom_0$, $\hmom_1$ and $\hmom_2$ are the constants 
from~\eqref{eq:hmomzero},~\eqref{eq:hmomone} and~\eqref{eq:dtwo} respectively,
and note that $|\para|\leq C B^{7/48}$ for some constant $C$.

Since $7/48<1/6$ we may apply Theorem~\ref{thm:goodtrial}(i). 
It follows by Proposition~\ref{prop:gap} that 
\begin{equation*}
\frac{1}{B}\eigone{\Ham_m(B)}=\lambda_0 + \lambda_2 B^{-2/6}+\lambda_3 B^{-3/6} 
+\lambda_4(\para)B^{-4/6}+\bigo((1+|\para|^2)B^{-5/6})
,
\end{equation*}
as $B\to\infty$.

For large $|\para|$ and $B$, we have
\begin{equation*}
\lambda_4(\para)B^{-4/6}+\bigo((1+|\para|^2)B^{-5/6})\geq \lambda_4+2
\end{equation*}
where $\lambda_4$ is the constant in~\eqref{eq:lfour}. 
We let $\k_{1/6}$ and $B_{1/6}$ correspond to the constants for which 
$\k_{1/6}B^{1/6}\leq |\mom_3|\leq \k_{5/16}B^{5/16}$ and $B>B_{1/6}$ implies 
that 
\begin{equation*}
\eigone{\Ham_m(B)}\geq \lambda_0 B + \lambda_2 B^{4/6}+\lambda_3 B^{3/6} 
+(\lambda_4+1)B^{2/6}.
\end{equation*}
This clearly implies~\eqref{eq:refinedlower}.

\subsubsection{Proof for $\k_{0}\leq |\mom_3| < \k_{1/6} B^{1/6}$, 
final step} 
Assume that $|\mom_3| < \k_{1/6} B^{1/6}$ and that $B>B_{1/6}$, with the 
constants $\k_{1/6}$ and $B_{1/6}$ from the previous step. Again, we let 
$\para$ be given by
\begin{equation*}
m = \frac{B}{2}+\hmom_0 B^{1/2}+\hmom_1B^{1/3}+(\hmom_2+\para)B^{1/6}.
\end{equation*}
This time $|\para|\leq C$, for some constant $C$, so we can apply 
Theorem~\ref{thm:goodtrial}(ii) combined with Proposition~\ref{prop:gap}. 
Recall that $\lambda_4(\para)=\dGdiff\para^2+\lambda_4$ and 
$\lambda_5(\para)=\lambda_5+a_1\para + a_2\para^2+a_3 \para^3$.
We rewrite the last two terms in the eigenvalue expansion from 
Theorem~\ref{thm:goodtrial}(ii) as
\begin{multline*}
\lambda_4(\para)B^{-4/6}+\lambda_5(\para)B^{-5/6} \\
= \lambda_4B^{-4/6}+\lambda_5B^{-5/6}
+\dGdiff\para^2 B^{-4/6}
+(a_1\para+a_2\para^2+ a_3\para^3)B^{-5/6}.
\end{multline*}
If we choose $\widetilde{C}$ sufficiently large
and $C\geq \widetilde{C}B^{-1/6}$ then the term $\dGdiff \para^2 B^{-4/6}$ will 
dominate both $(a_1\para+a_2\para^2+a_3\para^3)B^{-5/6}$ and the error which is
bounded by some constant times $B^{-6/6}$, indeed, we can get that all these 
three terms are bounded from below by $\frac{1}{2}\dGdiff C^2 B^{-6/6}$.

Therefore we find that there exist 
constants $B_0$ and $\k_{0}$ such that if it holds that
$\k_{0}\leq |\mom_3|\leq\k_{1/6}B^{1/6}$ and $B>B_0$ then
\begin{equation*}
\eigone{\Ham_m}(B)\geq \lambda_0B  + \lambda_2 B^{4/6}+\lambda_3 B^{3/6} 
+\lambda_4 B^{2/6}+\lambda_5 B^{1/6}+K.
\end{equation*}
This finishes the proof of Theorem~\ref{thm:refinedlower}.
\qed

%
%
%
%
%
%
%  PROOF OF MAIN THEOREM
%
%
%
%
%
%
%

\section{Proof of Theorem~\ref{thm:main}}\label{sec:proof}

For bounded $\mom_3$ we combine Theorem~\ref{thm:goodtrial}(iii) with 
Proposition~\ref{prop:gap} to get the asymptotic formula
\begin{equation}\label{eq:finest}
\eigone{\Ham_m(B)}=\lambda_0 B+ \lambda_2 B^{4/6}+\lambda_3 B^{3/6} 
+\lambda_4B^{2/6}+\lambda_5 B^{1/6}+\lambda_6(\mom_3)+
\bigo\bigl(B^{-1/6}\bigr)
\end{equation}
as $B\to\infty$.
Comparing the lower bound from~\eqref{eq:refinedlower} with~\eqref{eq:finest}, 
we find that the lowest eigenvalue is smallest for bounded $\mom_3$, and that
its asymptotic expansion then is given by~\eqref{eq:finest}. For bounded 
$\mom_3$ we see from~\eqref{eq:lsix} that the smallest value of 
$\lambda_6(\mom_3)$ is given for $\mom_3=\widehat{\mom}_3$. However, since $m$ 
must be an integer, we are not free to choose $\mom_3=\mom_3(m,B)$ arbitrarily. 
With 
\begin{equation*}
\Delta_B = \inf_{m\in\mathbb{Z}}\bigl|\mom_3(m,B)-\widehat{\mom}_3\bigr|
\end{equation*}
as in~\eqref{eq:deltab} we find that the smallest possible $\lambda_6(\mom_3)$ 
is given by
\begin{equation}\label{eq:lsixsec}
\lambda_6 = \dGdiff\Delta_B^2+\C.
\end{equation}
This finishes the proof of Theorem~\ref{thm:main}.\qed

%
%
%
%
%
%   Monotonicity
%
%
%
%
%
%
\section{Monotonicity of $\eigone{\Ham(B)}$, Proof of
Theorem~\ref{thm:monotonicity}}
\label{sec:monotonicity}

We first note that by perturbation theory it holds that
\begin{equation}
\label{eq:pertineq}
\eigonep{\Ham(B)}\leq \eigonem{\Ham(B)}
\end{equation}
for all $B>0$.

From Theorem~\ref{thm:main} we know that the lowest eigenvalue
$\eigone{\Ham(B)}$ of $\Ham(B)$ satisfies
\begin{equation}
\label{eq:monasymp}
\eigone{\Ham(B)}= \Theta_0 B + \lambda_2 B^{2/3}+\lambda_3 B^{1/2}+\lambda_4
B^{1/3}+\lambda_5 B^{1/6}+\dGdiff \Delta_B^2+\C +\bigo(B^{-1/6})
\end{equation}
where
\begin{equation*}
\Delta_B = \inf_{m\in\mathbb{Z}} \bigl|\mom_3(m,B)-\widehat{\mom}_3\bigr|
\end{equation*}
and $\mom_3(m,B) = m-\frac{B}{2}-\hmom_0\sqrt{B}-\hmom_1 B^{1/3}-\hmom_2
B^{1/6}$.

It is proved in~\cite{fohebook} that the derivatives $\eigonepm{\Ham(B)}$ 
satisfies
\begin{equation}\label{eq:liminf}
\begin{aligned}
\liminf_{B\to\infty} \eigonep{\Ham(B)} & \geq \limsup_{\epsilon \to 0+} 
\frac{1}{\epsilon} \liminf_{B\to\infty}
\bigl(\eigone{\Ham(B+\epsilon)}-\eigone{\Ham(B)}\bigr),\\
\limsup_{B\to\infty} \eigonem{\Ham(B)} & \leq \liminf_{\epsilon \to 0+} 
\frac{1}{\epsilon} \limsup_{B\to\infty}
\bigl(\eigone{\Ham(B)}-\eigone{\Ham(B-\epsilon)}\bigr).
\end{aligned}
\end{equation}
We start with the right derivative $\eigonep{\Ham(B)}$, and 
use~\eqref{eq:monasymp} to write
\begin{equation*}
\frac{\eigone{\Ham(B+\epsilon)}-\eigone{\Ham(B)}}{\epsilon} = \Theta_0 +
\frac{g(B+\epsilon)-g(B)}{\epsilon} +
\dGdiff\frac{\Delta_{B+\epsilon}^2-\Delta_B^2}{\epsilon}
+\frac{f(B+\epsilon)-f(B)}{\epsilon}
\end{equation*}
where
\begin{equation*}
g(B) = \lambda_2 B^{2/3}+\lambda_3 B^{1/2}+\lambda_4 B^{1/3}+\lambda_5
B^{1/6}+\C
\end{equation*}
and $f(B)$ is a function satisfying $\lim_{B\to\infty}f(B)=0$. For 
any fixed $\epsilon>0$ we clearly have
\begin{equation*}
\lim_{B\to\infty} g(B+\epsilon)-g(B)=0.
\end{equation*}
Consider the term involving
$\frac{\Delta_{B+\epsilon}^2-\Delta_B^2}{\epsilon}$. We note that there exist
integers $m_B$ and $m_{B+\epsilon}$ such that
\begin{equation*}
\begin{aligned}
\Delta_B &= \Bigl|m_B-\frac{B}{2}-\hmom_0\sqrt{B}-\hmom_1 B^{1/3}-\hmom_2
B^{1/6}-\hmom_3\Bigr|,\quad\text{and}\\
\Delta_{B+\epsilon} &=
\Bigl|m_{B+\epsilon}-\frac{B+\epsilon}{2}-\hmom_0\sqrt{B+\epsilon}-\hmom_1
(B+\epsilon)^{1/3}-\hmom_2 (B+\epsilon)^{1/6}-\hmom_3\Bigr|.
\end{aligned}
\end{equation*}
We note that
\begin{equation*}
\begin{aligned}
\Delta_B &= \inf_{m\in\mathbb{Z}}\Bigl|m-\frac{B}{2}-\hmom_0\sqrt{B}-\hmom_1
B^{1/3}-\hmom_2 B^{1/6}-\hmom_3\Bigr|\\
 & \leq \Bigl|m_{B+\epsilon}-\frac{B}{2}-\hmom_0\sqrt{B}-\hmom_1
 B^{1/3}-\hmom_2 B^{1/6}-\hmom_3\Bigr|.
\end{aligned}
\end{equation*}
Using this, and the fact that $0\leq \Delta_B\leq 1/2$ for all $B$, we get by
the triangle inequality
\begin{multline*}
\frac{\Delta_{B+\epsilon}^2-\Delta_{B}^2}{\epsilon}
=\frac{(\Delta_{B+\epsilon}-\Delta_{B})}{\epsilon}(\Delta_{B+\epsilon}
+\Delta_{B})\\
\geq-\biggl|\frac{1}{2}+\hmom_0\frac{\sqrt{B+\epsilon}-\sqrt{B}}{\epsilon}
+\hmom_1\frac{(B+\epsilon)^{1/3}-B^{1/3}}{\epsilon}
+\hmom_2\frac{(B+\epsilon)^{1/6}-B^{1/6}}{\epsilon}\biggr|.
\end{multline*}
The right-hand side tends to $-1/2$ as $B\to\infty$, so for any fixed 
$\epsilon>0$ we get
\begin{equation*}
\frac{1}{\epsilon} \liminf_{B\to\infty}(\Delta_{B+\epsilon}^2-\Delta_B^2) 
\geq -\frac12.
\end{equation*}
Inserting these calculations in~\eqref{eq:liminf} it follows that
\begin{equation*}
\liminf_{B\to\infty} \eigonep{\Ham(B)} \geq \Theta_0 - \frac12\dGdiff.
\end{equation*}
According to~\eqref{eq:konstineq} the right-hand side $
\Theta_0 - \frac12\dGdiff>0$. This
finishes the proof of~\eqref{eq:lpineq}. The same calculations 
give~\eqref{eq:lmineq} for the left-derivative $\eigonem{\Ham(B)}$. 
We conclude the proof of Theorem~\ref{thm:monotonicity} by noting
that the equations~\eqref{eq:lpineq} and~\eqref{eq:pertineq} imply that
$\eigone{\Ham(B)}$ is increasing for large $B$.\qed

%
%
%
%
%
%
%  Appendix
%
%
%
%
%
%
%

\begin{appendix}

%
%
%
%
%
%
%  Model operators
%
%
%
%
%
%
%

\section{Model operators}
\label{sec:modeloperators}

In this appendix, we consider two self-adjoint model operators. The first one 
is an operator in $L^2(\mathbb{R}_+)$ that was introduced by 
Saint-James and de~Gennes~\cite{jage}.
The second one is an operator in $L^2(\mathbb{R})$ first studied by Montgomery
in~\cite{mo}. 

\subsection{A general Lemma}
We start by giving a general lemma that will 
enable us to give moment formulas for the two operators under study.

\begin{lemma}
\label{lem:moment}
Let $-\infty\leq\alpha<\beta\leq\infty$, and $p\in C^1[\alpha,\beta]$. 
(If $\alpha=-\infty$ and $\beta=+\infty$ then we assume that 
$\lim_{x\to\pm\infty} p(x)=+\infty$). 
Assume that for some $\lambda\in\mathbb{R}$ and $u\in L^2(\alpha,\beta)$ it 
holds that
\begin{equation*}
-u''+pu=\lambda u\quad \text{for all } x\in[\alpha,\beta].
\end{equation*} 
Then, for any polynomial $b$, it holds that
\begin{equation}
\label{eq:xmoment}
\int_{\alpha}^\beta \bigl[b'''+4(\lambda -p)b' - 2p'b\bigr]u^2 \ed x = 
\big[2b(u')^2 +b''u^2-2b'uu'+2(\lambda-p)bu^2 \bigr]_\alpha^\beta.
\end{equation}
\end{lemma}

\begin{prf}
In the case $\alpha=-\infty$ and/or $\beta=+\infty$, the additional assumption
on $p$ implies that $u$ decays exponentially at $\alpha$ and/or $\beta$ (the 
proof is the same as in Lemma~\ref{lem:montgomeryschwartz}). 

One could use the same reasoning as in~\cite{best}. However, a simple 
calculation shows that the derivative of the expressions inside the brackets in
the right-hand side of~\eqref{eq:xmoment} equals the integrand on the left-hand
side.
\qed
\end{prf}

\subsection{The de Gennes operator}\label{sec:dG}
For $\xi\in\mathbb{R}$ we define the operator $\dG(\xi)$ as the self-adjoint 
Neumann extension in $L^2(\mathbb{R}_+)$, acting as
\begin{displaymath}
\begin{aligned}
(\dG(\xi) u)(x) &= -u''(x)+(x+\xi)^2u(x),\\
u'(0)&=0.
\end{aligned}
\end{displaymath}
Denote by $\eigone{\dG}(\xi)$ the lowest eigenvalue of $\dG(\xi)$ and
$\Theta_0=\inf_{\xi\in\mathbb{R}} \eigone{\dG}(\xi)$. We refer 
to~\cite{fohebook} for a discussion of the results summarized in the following
lemma.
\begin{lemma}
\label{lem:thetazero}
The function $\xi\mapsto \eigone{\dG}(\xi)$ is smooth. Moreover,
\begin{enumerate}
\item $\lim_{\xi\to+\infty}\eigone{\dG}(\xi)=+\infty$.
\item $\lim_{\xi\to-\infty}\eigone{\dG}(\xi)=1$.
\item The function $\eigone{\dG}(\xi)$ attains its minimum value $\Theta_0$,
$\frac12<\Theta_0<1$, at a unique point $\xi_0<0$.
\item $\eigone{\dG}(\xi)$ is decreasing for $\xi<\xi_0$ and increasing for
$\xi>\xi_0$.
\item The number $\dGdiff:=\frac{1}{2}\eigone{\dG}''(\xi_0)$ satisfies
$0<\dGdiff<1$.
\end{enumerate}
\end{lemma}
If we denote by $v_\xi$ the normalized eigenfunction of $\dG(\xi)$ corresponding 
to the eigenvalue $\eigone{\dG}(\xi)$, then we introduce the regularized 
resolvent
\begin{equation}
\label{eq:rreg}
\Rreg(\xi) g = 
\begin{cases}
(\dG(\xi)-\eigone{\dG}(\xi))^{-1}g, & g\perp v_\xi,\\
0 & g\parallel v_\xi,
\end{cases}
\end{equation}
and let $\Rreg=\Rreg(\xi_0)$ and $u_0=v_{\xi_0}$.
\begin{lemma}[\cite{fohe2}, Lemma~A.5]
\label{lem:degennesschwartz}
The function $u_0$ belongs to $\mathcal{S}(\mathbb{R}_+)$ and $\Rreg$ maps 
$\mathcal{S}(\mathbb{R}_+)$ continuously into itself.
\end{lemma}
\begin{lemma}[\cite{best}, equations~(2.34)--(2.36)]
\label{lem:best}
The following equalities hold
\begin{equation}
\label{eq:star}
\begin{aligned}
\int_0^\infty u_0^2 \ed x  &= 1,& \int_0^\infty (x+\xi_0)u_0^2 \ed x &= 0,\\
\int_0^\infty (x+\xi_0)^2 u_0^2 \ed x &= \frac{\Theta_0}{2},& \int_0^\infty
(x+\xi_0)^3 u_0^2 \ed x &= \frac{u_0^2(0)}{6}.
\end{aligned}
\end{equation}
\end{lemma}
We introduce the integrals
\begin{equation}\label{eq:kj}
k_j(\xi) = \int_0^\infty (x+\xi)u \bigl[\Rreg(\xi) (x+\xi)\bigr]^j u \ed x
\end{equation}
and $k_j=k_j(\xi_0)$.
\begin{lemma}[\cite{fohe2}, Proposition~A.3]
It holds that
\begin{equation}
\label{eq:respar}
\dGdiff = \frac{1}{2}\eigone{\dG}''(\xi_0)=1-4k_1.
\end{equation}
\end{lemma}

\begin{remark}
Numerical calculations of the constants $\xi_0$, $\Theta_0$, 
$\dGdiff$ and $u_0(0)$ has been carried out in~\cite{bonn1}.

We give a new approach. It is readily seen that the decaying normalized 
solution to $\dG(\xi)u=\eigone{\dG}(\xi)u$
is given by 
$u(x)=c e^{-\frac12(x+\xi)^2}H_{\frac{1}{2}(\eigone{\dG}(\xi)-1)}(x+\xi)$. Here
$c$ denotes a normalization constant and $H_\nu$ is the Hermite function, that 
solves $-y''(x)+2xy'(x)-2\nu y(x)=0$. The boundary 
condition $u'(0)=0$ transforms into
 $(\eigone{\dG}(\xi)-1)H_{\frac{1}{2}(\eigone{\dG}(\xi)-3)}(\xi)
-\xi H_{\frac{1}{2}(\eigone{\dG}(\xi)-1)}(\xi)=0$ and since 
$\Theta_0=\eigone{\dG}(\xi_0)=\xi_0^2$, we find that
$\xi_0$ should be the largest (it is negative!) number that solves
\begin{equation*}
(\xi^2-1)H_{\frac{1}{2}(\xi^2-3)}(\xi)
-\xi H_{\frac{1}{2}(\xi^2-1)}(\xi)=0.
\end{equation*}
Numerical calculations in Mathematica give
\begin{gather*}
\xi_0 \approx -0.76818365314,\quad \Theta_0\approx 0.59010612495,\\
u_0(0)\approx 0.87304313851,\quad \dGdiff \approx 0.58551290029.
\end{gather*}
\end{remark}

\subsection{The Montgomery operator}\label{sec:Mg}
Next, we turn to the Montgomery operator $\M(\mom)$, $\mom\in\mathbb{R}$,
defined as the self-adjoint operator in $L^2(\mathbb{R})$ acting as
\begin{equation*}
(\M(\mom) u)(\rho) = -u''(\rho)+(\mom+\rho^2)^2 u(\rho),
\quad -\infty < \rho < \infty
\end{equation*}
Denote by $\eigone{\M}(\mom)$ the lowest eigenvalue of $\M(\mom)$ and
$\hat{\nu}_0=\inf_{\mom\in\mathbb{R}} \eigone{\M}(\mom)$.

\begin{lemma}[\cite{mo,pakw,helf}]
\label{lem:montgomery}
The function $\mom\mapsto \eigone{\M}(\mom)$ is smooth and satisfies
$\lim_{\mom\to\pm\infty}\eigone{\M}(\mom)=+\infty$. Moreover, the minimal
value $\hat{\nu}_0>0$ of $\eigone{\M}(\mom)$ is attained
at a unique point $\hmom<0$, and $\eigone{\M}''(\hmom)>0$.
\end{lemma}

Let us denote by $\phi$ the eigenfunction corresponding to 
$\eigone{\M}(\widehat{\mom})$. It is known that such an eigenfunction belongs
to $C^\infty(\mathbb{R})$. We show that $\phi$ and its
derivatives decay exponentially, which implies that $\phi$ belongs to 
$\mathcal{S}(\mathbb{R})$.

\begin{lemma}\label{lem:montgomeryschwartz} 
Let $\phi$ be the ground state of $\M(\widehat{\mom})$.
For any $0<a<1/3$ and nonnegative integer $k$ there exist a constant 
$C_k$ such that
\begin{equation}
\int_{\mathbb{R}} e^{2a|\rho|^3}
\Bigl(|\phi|^2+|\rho^k\phi|^2+|(\PD{}{\rho})^k\phi|^2\Bigr)\ed \rho 
\leq  C_k \int_{\mathbb{R}} |\phi|^2\ed \rho.
\end{equation}
\end{lemma}

\begin{prf}
Let $a<1/3$ be given. For $\epsilon>0$ we define 
$v_\epsilon(\rho)= \bigl(|\rho|/(1+\epsilon |\rho|)\bigr)^3$. Then,
for fixed $\rho$, $v_\epsilon(\rho)$ is monotonically increasing to $|\rho|^3$ 
as $\epsilon\to 0$. Moreover it holds that
\begin{equation}\label{eq:vepsest}
|v_\epsilon'(\rho)|\leq 3|\rho|^2
\end{equation}
for all $\epsilon>0$ and all $\rho\in\mathbb{R}$. We let 
$\chi_{1,M}$ and $\chi_{2,M}$ denote the same functions as in the proof of
Lemma~\ref{lem:rholocthree}. The IMS formula gives
\begin{multline}\label{eq:montfirst}
\int_{\mathbb{R}} \bigl|\PD{}{\rho}(\chi_{2,M}e^{a v_\epsilon}\phi)\bigr|^2
+ \Bigl(\bigl(\widehat{\mom}+\rho^2\bigr)^2-
\eigone{\M}(\widehat{\mom})\Bigr)|\chi_{2,M}e^{av_\epsilon}\phi|^2 \ed \rho\\
=\int_{\mathbb{R}} \bigl|\PD{}{\rho}(\chi_{2,M}e^{a v_\epsilon})\phi\bigr|^2 
\ed \rho 
\end{multline}
We use the Cauchy-Schwarz inequality on the right-hand side, to get, 
for any $\cpar>0$,
\begin{multline}\label{eq:montsec}
\int_{\mathbb{R}} \bigl|\PD{}{\rho}(\chi_{2,M}e^{a v_\epsilon})\phi\bigr|^2 
\ed \rho \\
\leq(1+\cpar)\int_{\mathbb{R}} \bigl|a v_\epsilon'e^{a v_\epsilon}\chi_{2,M}
\phi\bigr|^2 \ed \rho 
+ \bigl(1+\frac{1}{\cpar}\bigr)\int_{\mathbb{R}} \bigl|(\PD{}{\rho}\chi_{2,M})
e^{a v_\epsilon}\phi\bigr|^2 \ed \rho 
\end{multline}
We recall that $|\PD{}{\rho}\chi_{2,M}|\leq \ccon_2/M$ for all $\rho$ and that 
$\PD{}{\rho}\chi_{2,M}$ has support in the set 
$\{\rho\in\mathbb{R}\mid M\leq \rho\leq 2M\}$. We 
implement \eqref{eq:vepsest} and~\eqref{eq:montsec} in~\eqref{eq:montfirst} to 
find that
\begin{multline}\label{eq:montthird}
\int_{\mathbb{R}} \bigl|\PD{}{\rho}(\chi_{2,M}e^{a v_\epsilon}\phi)\bigr|^2
+ \Bigl(\bigl(\widehat{\mom}+\rho^2\bigr)^2-
\eigone{\M}(\widehat{\mom})-(1+\cpar)9a^2|\rho|^4\Bigr)
|\chi_{2,M}e^{av_\epsilon}\phi|^2 \ed \rho\\
\leq \bigl(1+\frac{1}{\cpar}\bigr)\frac{\ccon_1^2}{M^2}e^{a(2M)^3}
\int_{\mathbb{R}} |\phi|^2 \ed \rho 
\end{multline}
Since $a<1/3$ we can choose $\cpar$ so small that $(1+\cpar)9a^2<1$. With this
choice of $\cpar$ we can find $M$ so large that 
\begin{equation*}
\Bigl(\bigl(\widehat{\mom}+\rho^2\bigr)^2-
\eigone{\M}(\widehat{\mom})-(1+\cpar)9a^2|\rho|^4\Bigr)\geq 1
\end{equation*}
for all $\rho$ on the support of $\chi_{2,M}$. This together with the trivial
bound for small $\rho$ settles the result for $\phi$. We might also use the
first term in~\eqref{eq:montthird} to prove the result for $\PD{}{\rho}$. The 
statement for $\rho^k\phi$ follows from this by decreasing $a$ 
(or to be more precise, prove the result for $\phi$ for $1/3>a'>a$ and then
decrease this $a'$ to $a$). The result for higher derivatives is now
a consequence of induction, using the eigenvalue equation.\qed
\end{prf}

Let us define the regularized resolvent $\tRreg$ as
\begin{equation}
\label{eq:trreg}
\tRreg u = 
\begin{cases}
(\M(\widehat{\mom})-\eigone{\M}(\widehat{\mom}))^{-1}u, & u\perp \phi,\\
0 & u\parallel \phi.
\end{cases}
\end{equation}

We show that if $u$ and its derivatives decay exponentially, then the same
is true for $\tRreg u$.

\begin{lemma}\label{lem:rhoexpdec}
Let $0<a<1/3$. Assume that  $u$, $\rho^k u$ and $(\PD{}{\rho})^k u$ belong to 
$L^2(\mathbb{R},e^{2a|\rho|^3}\ed \rho)$ for all non-negative integers $k$. 
Then, for any $b<a$, $\tRreg u$,  $\rho^l\tRreg u$ and 
$(\PD{}{\rho}^l) \tRreg u$ belong to $L^2(\mathbb{R},e^{2b|\rho|^3}\ed \rho)$ 
for all non-negative integers $l$.
\end{lemma}

\begin{prf}
Let $w=\tRreg u$, so that 
\begin{equation}\label{eq:resolveq}
-\PD{}{\rho}^2 w + (\widehat{\mom}+\rho^2)^2w-\eigone{\M}(\widehat{\mom})w = u, 
\end{equation}
with $u$ 
as in the assumptions. Let $\epsilon>0$ and let $v_\epsilon$ be the same 
function as in the proof of the previous lemma, 
$v_\epsilon(\rho)=\bigl(|\rho|/(1+\epsilon|\rho|)\bigr)^3$. 
We also let $\chi_{1,M}$ and $\chi_{2,M}$ be the same cut-off functions as in 
the proof of Lemma~\ref{lem:rholocthree}. An integration by parts gives
\begin{multline*}
\int_{\mathbb{R}} 
\bigl|\PD{}{\rho}\bigl(\chi_{2,M}e^{a v_\epsilon} w\bigr)\bigr|^2
+ \bigl|\bigr(\widehat{\mom}+\rho^2\bigr)\chi_{2,M}e^{a v_\epsilon} w\bigr|^2 
\ed \rho\\
=\eigone{\M}(\widehat{\mom})\int_{\mathbb{R}}
\bigl|\chi_{2,M}e^{av_\epsilon} w|^2
\ed \rho
+\int_{\mathbb{R}} \bigl|\PD{}{\rho}(\chi_{2,M}e^{a v_\epsilon})w\bigr|^2 
\ed \rho
+\int_{\mathbb{R}} \bigl|\chi_{2,M}e^{a v_\epsilon}\bigr|^2 w  u \ed \rho.
\end{multline*}
We use the Cauchy-Schwarz inequality for the last term and move terms to the 
left-hand side,
\begin{equation*}
\int_{\mathbb{R}} 
\bigl|\PD{}{\rho}\bigl(\chi_{2,M}e^{a v_\epsilon} w\bigr)\bigr|^2
+ P(\rho)\bigl|e^{a v_\epsilon} w\bigr|^2 \ed \rho
\leq 2 \int_{\mathbb{R}}\bigl|\chi_{2,M}e^{av_\epsilon} u|^2 \ed \rho,
\end{equation*}
where
\begin{equation*}
P(\rho)=\Bigl(\bigl|\bigr(\widehat{\mom}+\rho^2\bigr)\chi_{2,M}\bigr|^2-
\bigl(\eigone{\M}(\widehat{\mom})-2\bigr)|\chi_{2,M}|^2 - 
\bigl|\PD{}{\rho}(\chi_{2,M})+av_\epsilon'\chi_{2,M}\bigr|^2  \Bigr).
\end{equation*}
By~\eqref{eq:vepsest}, the first term in $P$ is dominant for large $\rho$, 
so if we choose $M$ large
enough we have $P(\rho)\geq 1$ for all $\rho$ on the support of $\chi_{2,M}$, 
and thus we get, for such $M$, that
\begin{equation}\label{eq:helpderiv}
\int_{\{|\rho|>M\}} 
\bigl|\PD{}{\rho}\bigl(\chi_{2,M}e^{a v_\epsilon} w\bigr)\bigr|^2
+ |e^{a v_\epsilon} w|^2 \ed \rho
\leq 2 \int_{\mathbb{R}}\bigl|\chi_{2,M}e^{av_\epsilon} u|^2 \ed \rho.
\end{equation}
The right-hand side is clearly bounded by 
$2 \int_{\mathbb{R}}|u|^2 e^{2a |\rho|^3} \ed \rho$ which is bounded by
assumption. We let $\epsilon\to 0$ and use monotone convergence to conclude
\begin{equation*}
\int_{\{|\rho|>M\}} |w|^2 e^{2 a |\rho|^3}\ed \rho
\leq 2 \int_{\mathbb{R}}|u|^2 e^{2a|\rho|^3}\ed \rho.
\end{equation*}
The estimate for $|\rho|<M$ is trivial. This proves the statement in our lemma
for $w$ (with $b=a$). The estimate for $\rho^l w$ is simple 
if we just decrease $a$ to $b$, so that $\rho^l$ is dominated by the exponential
$\exp(2(b-a)|\rho|^3)$ for large $\rho$. 
For the first derivative $\PD{}{\rho}w$ we might use~\eqref{eq:helpderiv} and 
for higher derivatives we continue by induction, using~\eqref{eq:resolveq}.
\qed
\end{prf}

We will several times encounter a variant of the Montgomery operator. For $k>0$,
we denote by $\widetilde{\M}(\mom)$ the operator
\begin{equation}
\label{eq:ourmontgomery}
(\widetilde{\M}(\mom) u)(\rho) =
-u''(\rho)+k\Bigl(\mom+\frac{\rho^2}{2}\Bigr)^2 u(\rho),
\quad -\infty < \rho < \infty
\end{equation}
A change of coordinates
\begin{equation*}
\tilde{\rho}=k^{1/6}2^{-1/3}\rho
\end{equation*}
transforms $\widetilde{\M}(\mom)$ into
\begin{equation*}
\widetilde{\M}(\mom)=k^{1/3}2^{-2/3}\M(k^{1/3}2^{1/3}\mom)
\end{equation*}
and so for the lowest eigenvalue it holds that
\begin{equation}
\label{eq:montgomerytwo}
\eigone{\widetilde{\M}}(\mom)
=k^{1/3}2^{-2/3}\eigone{\M}(k^{1/3}2^{1/3}\mom).
\end{equation}
In the case when $k=\dGdiff$ we write
\begin{equation*}
\tilde{\mom}=(2\dGdiff)^{-1/3}\widehat{\mom}
\end{equation*}
and we get that $\eigone{\widetilde{\M}}(\mom)$ is minimal for
$\mom=\tilde{\mom}$ and that
\begin{equation}
\label{eq:montgomerythree}
\eigone{\widetilde{\M}}(\tilde\mom)
=2^{-2/3}\dGdiff^{1/3}\eigone{\M}(\hmom) 
=2^{-2/3}\dGdiff^{1/3}\hat{\nu}_0
=\widehat{\gamma}_0.
\end{equation}
For the case $k=\dGdiff$ we also introduce the moments
\begin{equation*}
M^l_{j,k} = \int_{-\infty}^\infty 
\Bigl(\tilde{\mom}+\frac{\rho^2}{2}\Bigr)^l \phi_j \phi_k \ed\rho.
\end{equation*}
Here $\phi_0$ is the first normalized eigenfunction of 
$\widetilde{\M}(\tilde{\mom})$ for
$k=\dGdiff$, and $\phi_j$, $j\geq 1$ are constructed via the Gru{\v{s}}in 
method in Section~\ref{sec:upper}. 

\begin{lemma}
It holds that
\begin{equation*}
M_{0,0}^0=1,\quad M_{0,0}^1 = 0,\quad 
M_{0,0}^2  = \frac{\hat{\nu}_0}{3(2\dGdiff)^{2/3}},\quad \text{and}\quad 
M_{0,0}^3  = \frac{1}{6\dGdiff}
            -\frac{\tilde{\mom}\hat{\nu}_0}{3(2\dGdiff)^{2/3}}.
\end{equation*}
\end{lemma}

\begin{prf}
The first formula is just the normalization of $\phi_0$ and the second one 
follows by a perturbation argument, just as for the de~Gennes model.

We use Lemma~\ref{lem:moment} to calculate $M_{0,0}^l$ for $l\geq 2$. Indeed, 
with $p(\rho)=\bigl(\tilde{\mom}+\frac{\rho^2}{2}\bigr)^2$ and 
$\lambda=\widehat{\gamma}_0$, the formula~\eqref{eq:xmoment} 
becomes
\begin{equation}
\label{eq:rhomomentmont}
\int_{-\infty}^\infty \biggl[b'''+4\biggl(\widehat{\gamma}_0
- \Bigl(\tilde{\mom}+\frac{\rho^2}{2}\Bigr)^2\biggr)b' 
- 2\rho \Bigl(\tilde{\mom}+\frac{\rho^2}{2}\Bigr)b\biggr]\phi_0^2 \ed\rho = 0.
\end{equation}
The choice $b(\rho)=\rho^{2l-1}$ gives $M_{0,0}^{l+1}$ provided that 
the previous moment formulas are known. 
Especially the choices $b(\rho)=\rho$ and $b(\rho)=\rho^3$ give the announced 
formulas for $M_{0,0}^2$ and $M_{0,0}^3$.\qed
\end{prf}

\end{appendix}

\subsubsection*{Acknowledgements} The authors thank Ayman Kachmar and
Bernard Helffer for 
fruitful discussions. Both authors were supported by the Lundbeck Foundation. 
SF is also supported by the Danish Research Council and by the European 
Research Council under the European Community's Seventh Framework 
Program (FP7/2007--2013)/ERC grant agreement 202859.

\bibliographystyle{abbrv}
\def\cprime{$'$}

\end{document}